\RequirePackage{ifpdf}
\ifpdf 
\documentclass[pdftex]{sigma}
\else
\documentclass{sigma}
\fi

\def\vecp{\mbox{$\hat{\vec{p}}$} \,{}'_{\!a}}
\def\vecpp{\mbox{$\hat{\vec{p}}$}_{a}}

\begin{document}

\allowdisplaybreaks

\renewcommand{\PaperNumber}{117}

\FirstPageHeading

\renewcommand{\thefootnote}{$\star$}

\ShortArticleName{Quasi-Particle Excitation in QED}

\ArticleName{Self-Localized Quasi-Particle Excitation in Quantum
Electrodynamics and Its Physical Interpretation\footnote{This
paper is a contribution to the Proceedings of the Seventh
International Conference ``Symmetry in Nonlinear Mathematical
Physics'' (June 24--30, 2007, Kyiv, Ukraine). The full collection
is available at
\href{http://www.emis.de/journals/SIGMA/symmetry2007.html}{http://www.emis.de/journals/SIGMA/symmetry2007.html}}}

\Author{Ilya D. FERANCHUK and Sergey I. FERANCHUK}

\AuthorNameForHeading{I.D. Feranchuk and S.I. Feranchuk}

\Address{Department of Physics, Belarusian University, 4
Nezavisimosti  Ave., 220030, Minsk, Belarus }
\Email{\href{mailto:fer@open.by}{fer@open.by},
\href{mailto:sergey@feranchuk.linux.by}{sergey@feranchuk.linux.by}}

 \URLaddress{\url{http://www.theorphysics.bsu.by/Stuff/Feranchuk.htm},\newline
\null \, \  \ \ \ \url{http://sergey.feranchuk.net}}

\ArticleDates{Received October 21, 2007, in f\/inal form November
29, 2007; Published online December 07, 2007}

\Abstract{The self-localized quasi-particle excitation of the
electron-positron f\/ield (EPF) is found for the f\/irst time in
the framework of a standard form of the quantum electrodynamics.
This state is interpreted as the ``physical'' electron (positron)
and it allows one to solve the following problems: i) to express
the ``primary'' charge $e_0$ and the mass $m_0$ of the ``bare''
electron in terms of the observed values of $e$ and $m$ of the
``physical'' electron without any inf\/inite parameters and by
essentially nonperturbative way; ii) to consider $\mu$-meson as
another self-localized EPF state   and to estimate the ratio
$m_{\mu}/m$; iii) to prove that the self-localized state is
Lorentz-invariant and its energy spectrum corresponds to the
relativistic free particle with the observed mass  $m$; iv) to
show that the expansion in a power of the observed charge $e \ll
1$ corresponds to the strong coupling expansion in a power of the
``primary'' charge $e^{-1}_0 \sim e $ when the interaction between
the ``physical'' electron and the transverse electromagnetic
f\/ield is considered by means of the perturbation theory and all
terms of this series are free from the ultraviolet divergence. }


\Classification{81V05; 81V10; 83C47}

\section{Introduction}

It is no doubt at present that the Standard Model is the
fundamental basis for the theory of the electro-weak interaction
\cite{Weinberg}. It means that the quantum electrodynamics (QED)
is actually the part of the general gauge theory. Nevertheless,
QED   considered by itself as the isolated system remains the most
successful quantum f\/ield model that allows one to calculate the
observed characteristics of the electromagnetic processes with a
unique accuracy (for example, \cite{Akhiezer,Scharf95}).  It is
well known that these calculations are based on the series of
rules connected with the perturbation theory in the observed
charge $e$ of the ``physical'' electron and the renormalization
property of QED. The latter one means that the ``primary''
parameters of the theory (the charge $e_0$ and the mass $m_0$ of
the ``bare'' electron) that are def\/ined by the divergent
integrals, can be excluded from the observed values. However, even
the creators of the present form of QED were not satisf\/ied
because ``the calculation rules of QED are badly adjusted with the
logical foundations of quantum mechanics and they cannot be
considered as the satisfactory solution of the dif\/f\/iculties''
\cite[\S~81]{Dirac} and ``it is simply a way to sweep the
dif\/f\/iculties under the rug'' \cite{Feynman}.

There are a number motivations for      calculation of the
``bare'' electron characteristics $e_0$ and~$m_0$ in spite the
fact that these values are unobserved. First of all it is   the
question whether the system of Maxwell and Dirac equations as the
mathematical model for the quantum f\/ield system is closed and
self-consistent when considering the processes related to  the
interaction of electrons, positrons and photons?  In that sense
QED in the existing form of renormalization is an unclosed theory
because it includes an additional, external dimensional parameter
which enables the regularization of integrals. The contradiction
between the small value of the coupling constant and its
inf\/inite calculated value shows a logical inconsistency of the
perturbation theory in QED, so called ``Landau pole''
\cite[\S~128]{Lifshitz}. It is very important to consider the
renormalization problem in the framework of the ``logical
principles of quantum mechanics''   in order to develop some new
nonperturbative approach in the f\/ield theory in application to
the real physical system with completely def\/ined Hamiltonian. At
present nonperturbative methods are mainly studied for quite
abstract quantum f\/ield models with a strong interaction (for
example, \cite{Kleinert}). Such methods may be especially
interesting for the non-renormalized quantum f\/ield models.

It is also very essential that the dynamical description of the
internal structure of the ``physical'' electron gives the
fundamental possibility to consider  $\mu$-meson as an excited
state of the electron-positron f\/ield as it has been shown by
Dirac~\cite{Dirac1}.

The relation between the ``primary'' coupling constant  $e_0$ and
the charge $e$ is undetermined in the present form of QED.
Therefore it is possible that the value $e_0$ is large in spite
the observed renormalized charge being small   $e \ll 1$.
Precisely this possibility ($ e_0> 1$, but $e \ll 1$) is
investigated in the present paper in order to f\/ind a spectrum of
the quasi-particle excitations in QED without the perturbation
theory. Our main goal is to f\/ind such a form of the
renormalization   that would be logically consistent but the
calculation possibilities of QED for the observed values would be
preserved.

It is important to stress that the canonical QED model is
considered with a nonzero mass of the ``bare'' electron $m_0 \neq
0$ and without the chiral symmetry. This approach is distinguished
essentially from the nonperturbative analysis of the strong
coupling QED model with a zero mass of the fermion f\/ield (for
example, \cite{strong,strongSol} and citation therein). In the
latter case the observed particle mass  appears as the result of
spontaneous breaking of the chiral symmetry, however, the
cut-of\/f momentum $L$ is another undef\/ined parameter in such
theories.

The article is structured in the following way. In
Section~\ref{sec2} it is shown for the f\/irst time that the
self-localized one-particle excitation can be found in the
spectrum of QED Hamiltonian. This state cannot be calculated by
means of the perturbation theory as a power series of the coupling
constant $e_0$. The stability of this state is conditioned by a
localized charge distribution of the electron-positron f\/ield
coupled with a scalar component of the electromagnetic f\/ield.
The system of nonlinear equations for these spatially localized
distributions is derived and its numerical solution is obtained.

In Section~\ref{sec3} the self-localized excitation is interpreted
as the ``physical'' electron with the observed values of the
charge $e$ and mass $m$. It allowed us to express characteristics
$e_0$ and $m_0$ of the ``bare'' electron that are actually the
parameters of the initial Hamiltonian in terms of $e$ and $m$. It
is shown that the relation between these values includes the
singularity in the limit of $e \rightarrow 0$ and cannot be
calculated by means of the perturbation theory. This result cannot
be also obtained  in the framework of the ``quenched QED'' model
based on the Schwinger--Dyson equation \cite{miransky} or on the
variational approach \cite{varQED} because the charge
renormalization did not take into account in this model.

The considered physical interpretation of the self-localized state
leads also to an  important consequence. It is shown in
Section~\ref{sec3} that there is another one-particle excitation
of the electron-positron f\/ield with the same charge and spin as
for electron but with the larger mass. Following  Dirac
\cite{Dirac1} this excitation was considered as the ``physical''
$\mu$-meson and the ratio $m/m_{\mu}$ is calculated. The
calculated mass of $\mu$-meson proved to be very close to its
experimental value. It is essential that unlike  \cite{Dirac1} the
$\mu$-meson mass is calculated without any additional parameters
of the model with the exception of $e_0$ and $m_0$.

The localized charge distribution in the ``physical'' electron
corresponds to the spontaneous Lorentz symmetry breaking for the
considered system. This phenomenon is typical for the quantum
f\/ield theories with the particle-f\/ield strong coupling (for
example, the ``polaron'' problem~\cite{Pekar,Bogol}). It is shown
in Section~\ref{sec4} that  reconstruction of this symmetry leads
to the dependence of energy of the excitation   on its total
momentum. It is essential that this dependence corresponds exactly
the relativistic kinematics of a free particle with the observed
mass~$m$.

It is shown in Section~\ref{sec5} that the strong coupling series
in the initial QED Hamiltonian corresponds to the perturbation
theory in terms of observed charge   $e  \sim e_0^{-1} \ll 1 $.
However, all high-order corrections of this perturbation theory is
def\/ined by the convergent integrals without any additional
cut-of\/f parameters. The correspondence of these results with the
standard renormalization procedure when calculating the observed
characteristics of the electromagnetic processes is also
discussed.

\section{Self-localized state with zero momentum}\label{sec2}

It is well known that the spatially localized states are very
important for a lot of quantum f\/ield models. Let us remind the
one-dimensional model for the scalar f\/ield with the Hamiltonian
(for example, \cite{soliton}):
\begin{gather*}
\hat H = \int d x   \frac{1}{2}\left\{ \hat \pi^2   +
\left(\frac{\partial \hat \varphi(x)}{\partial x}\right)^2 +
\frac{1}{2}\lambda \left(\hat\varphi(x)^2 -
\frac{m^2}{\lambda}\right)^2\right\},
\end{gather*}
with the f\/ield operators  $\hat \varphi(x)$ and the
corresponding momentum operators $\hat \pi$.

This Hamiltonian has the eigenvector $ |\varphi\rangle$  with the
energy density localized in the vicinity of an arbitrary point
$x_0$. With the zero total momentum of the system the energy
density is distributed as follows:
\begin{gather}
\label{2} E(x) = \frac{m^4}{2 \lambda} \cosh^{-4}
\left[\frac{m}{\sqrt{2}}(x - x_0)\right],
\end{gather}
and corresponds to the f\/inite total energy  $E_0 =
(2\sqrt{2}/3)(m^3/\lambda)$. Lorentz-invariance of the system
leads to the relativistic dependence of the energy on the momentum
of the localized excitation~\cite{soliton}.

It is important for the further discussion that the state
(\ref{2}) cannot be derived by means of the perturbation theory
based on the coupling constant $\lambda$ connected with the
nonlinear interaction. One should separate the non zero classical
component from the f\/ield operators $\varphi(x) =
\langle\varphi|\hat \varphi|\varphi\rangle$ averaged over the
considered state even in the zeroth-order approximation. And the
classical function $\varphi(x)$ is not reduced to the plane wave
as in the zeroth-order perturbation theory. It is satisf\/ied to
the nonlinear dif\/ferential equation that is def\/ined by the
variation of the classical functional.

The Fr\"ohlich model for ``polaron'' problem gives another example
\cite{frohlich}. This model corresponds to the electron-phonon
interaction in the ionic crystals and is described by the
following Hamiltonian:
\begin{gather*}
\hat H = \frac{1}{2}\hat p^2 + \sum_{\vec k} a^+_{\vec k}a_{\vec
k} + 2^{3/4}\sqrt{\frac{\pi \alpha}{\Omega}}\sum_{\vec
k}\frac{1}{k} (a_{\vec k}+ a^+_{-\vec k})e^{i \vec k \cdot \vec
r}.
\end{gather*}

In this case the spatially localized electron state (``polaron'')
cannot be found also by means of the perturbation theory in terms
of the electron-phonon coupling constant $\alpha$.  The
variational wave function of the electron and the classical part
of the phonon f\/ield $u_{\vec k}= \langle a_{\vec k}\rangle $
should be calculated even in the zeroth-order approximation. The
translational symmetry of the initial  Hamiltonian leads to the
dependence of the energy of this one-particle excitation on its
total momentum~\cite{Bogol}.

Let us now consider the nonperturbative analysis of the spectrum
of the one-particle excitations of the QED Hamiltonian that is
def\/ined by the following form  (for example, \cite{Heitler}):
\begin{gather}
 \hat H = \int d \vec{r}:\! \{ \hat \psi^* (\vec{r}) [ \vec \alpha (\vec{p} + e_0 \hat{ \vec{ A} }
(\vec{r})) + \beta m_0] \hat \psi (\vec{r}) +  e_0\hat \varphi
(\vec{r}) \hat \rho (\vec{r}) - \frac{1}{2} ( \vec{\nabla}\hat
\varphi
(\vec{r}))^2\} \!: + \sum_{\vec k \lambda} \omega(\vec k) \hat n_{\vec k \lambda},\nonumber\\
\hat \rho (\vec{r}) = \frac{1}{2} [\hat \psi^* (\vec{r})\hat \psi
(\vec{r}) - \hat \psi (\vec{r})\hat \psi^*
(\vec{r})].\label{Eqn18}
\end{gather}

We suppose here that the f\/ield operators are given in the
Schr\"odinger  representation, the spinor components of the
electron-positron operators being def\/ined in the standard way
\cite{Heitler}
\begin{gather*}
 \hat \psi_{\nu} (\vec{r}) = \sum_{s} \int \frac{d\vec{p}}{(2\pi)^{3/2}} \{a_{\vec{p}s}
u_{\vec{p}s\nu} e^{i\vec{p}\vec{r}} +
b^+_{\vec{p}s} v_{-\vec{p}-s\nu} e^{-i\vec{p}\vec{r}}\},\nonumber\\
\hat \psi^*_{\nu} (\vec{r}) = \sum_{s} \int
\frac{d\vec{p}}{(2\pi)^{3/2}} \{a^+_{\vec{p}s} u^*_{\vec{p}s\nu}
e^{-i\vec{p}\vec{r}} + b_{\vec{p}s} v^*_{-\vec{p}-s\nu} e^{i\vec{p}\vec{r}}\}.
\end{gather*}

In these formulas $\hbar = c = 1$; the primary charge $(-e_0)$,
$e_0 >0$ and $m_0$ are considered as the parameters of the model;
the symbol $:\! \hat H \!:$ means the normal ordering of the
operators excluding the vacuum energy \cite{Dirac};
$\vec{\alpha}$, $\beta$ are   Dirac matrices; $u_{\vec{p}s\nu}$
and $v_{\vec{p}s\nu}$ are the components of the bispinors
corresponding to the solutions of Dirac equation for the free
``bare'' electron and positron with the momentum $\vec{p}$ and
spin $s$; $a_{\vec{p}s}(a^+_{\vec{p}s})$ and
$b_{\vec{p}s}(b^+_{\vec{p}s})$ are  the annihilation (creation)
operators for the ``bare'' electrons and positrons in the
corresponding states.  The f\/ield operator~$\hat{
\vec{A}}(\vec{r})$ and the operator of the photon number $\hat
n_{\vec k \lambda}$ are related to the transversal electromagnetic
f\/ield and their explicit form will be written below.

This Hamiltonian corresponds to the Coulomb gauge~\cite{Heitler},
when the electron-positron f\/ield interacts with the scalar
f\/ield and with the transversal photons of the electromagnetic
f\/ield. It will be shown below that the reconstruction of the QED
gauge symmetry connected with the longitudinal f\/ield does not
change the form of the considered one-particle excitation. In the
Coulomb gauge the operators of the scalar f\/ield
\begin{gather*}
\hat \varphi (\vec{r}) = \sqrt{4\pi}\int d\vec{k} \hat
\varphi_{\vec{k}}e^{i\vec{k}\vec{r}}
\end{gather*}
can be excluded from the Hamiltonian \cite{Heitler}. For that
purpose one should use the solution of the operator equations of
motion for   $\hat \varphi_{\vec{k}}$ assuming that the ``bare''
electrons are point-like particles and  ``self-action'' is
equivalent to the substitution of the initial mass for the
renormalized one. As a result the terms with scalar f\/ields in
the Hamiltonian are reduced to the Coulomb interaction between the
charged particles. However, this transformation of the
Hamiltonian~(\ref{Eqn18}) can not be used in this paper because
only the dynamics of the mass renormalization is the subject under
investigation.

There is another problem connected with a negative sign of the
term corresponding to the self-energy of the scalar f\/ield. If
the non-relativistic problems were considered then the operator of
the particle kinetic energy would be positively def\/ined and the
negative operator with the square-law dependence on  $\hat\varphi
(\vec{r})$ would lead to the ``fall on the center'' \cite{Landau}
as the energy minimum would be reached at an inf\/initely large
f\/ield amplitude. However, if the relativistic fermion  f\/ield
is considered then the operator of the free particle energy (the
f\/irst term in formula (\ref{Eqn18})) is not positively
def\/ined. Besides, the states of the system  with the negative
energy are f\/illed. Therefore, the stable state of the system
corresponds to the energy extremum(!) (the minimum one for
electron and the maximum one for positron excited states). It can
be reached at the f\/inite value of the f\/ield amplitude (see
below). The same reasons enable one to successfully use the states
with indef\/inite metric \cite{Akhiezer} in QED although it leads
to some dif\/f\/iculties in the non-relativistic quantum
mechanics.

According to our main assumption about the large value of the
initial coupling constant  $e_0$ we are to realize the
nonperturbative description of the excited state which is the
nearest to the vacuum state of the system. The basic method for
the nonperturbative estimation of the energy is the variational
approach with some trial state vector $|\Phi_1 \rangle$ for the
approximate description of the one-particle excitation. The
qualitative properties of the self-consistent excitation in the
strong coupling limit \cite{Pekar} show that such trial vector
should correspond to the general form of the wave packet formed by
the one-particle excitations of the ``bare'' electron-positron
f\/ield. Besides, the ef\/fect of polarization and the appearance
of the electrostatic f\/ield $\varphi (\vec{r})$ should be taken
into account, so we consider  $|\Phi_1\rangle $ to be the
eigenvector for the operator of the scalar f\/ield. Now, let us
introduce the following trial vector depending on the set of
variational classical functions $U_{\vec q s}$, $V_{\vec q s}$,
$\varphi (\vec r)$ for the approximate description of the
quasi-particle excited state of the system:
\begin{gather}
 |\Phi_1 \rangle \simeq |\Phi^{(0)}_1(U_{\vec q s}; V_{\vec q s};\varphi (\vec r) )\rangle = \int d \vec{q} \{
U_{\vec{q}s} a^+_{\vec{q} s} + V_{\vec{q}s} b^+_{\vec{q} s} \} |
0;
0;\varphi(\vec r)\rangle, \nonumber\\
\hat \varphi(\vec r)| 0; 0;\varphi(\vec r)\rangle =\varphi(\vec
r)| 0; 0;\varphi(\vec r)\rangle, \qquad a_{\vec{q} s}| 0;
0;\varphi(\vec r)\rangle = b_{\vec{q} s}| 0; 0;\varphi(\vec
r)\rangle = 0.\label{Eqn20}
\end{gather}

The ground state of the system is   $|\Phi_0\rangle =
|0;0;0\rangle$, if we use the same notation. It corresponds to the
vacuum of both interacting f\/ields.

Firstly, let us consider the excitation with the zero total
momentum. Then the constructed trial vector should satisfy the
normalized conditions resulting from the def\/inition of the total
momentum $\vec{P}$ and the observed charge $e$ of the ``physical''
particle:
\begin{gather}
\langle \Phi^{(0)}_1|\hat{ \vec {P}}|\Phi^{(0)}_1\rangle  =
\sum_{s} d \vec{q} \vec{q}\, [|U_{\vec{q}s}|^2 + |V_{\vec{q}s}|^2]
= \vec{P} = 0,\qquad \sum_{s} d \vec{q}\, [|U_{qs}|^2 +
|V_{qs}|^2] = 1,\label{Eqn21}
\\
\label{Eqn21a} \langle \Phi^{(0)}_1|\hat Q|\Phi^{(0)}_1\rangle =
e_0 \sum_{s} d \vec{q}\, [|V_{qs}|^2 - |U_{qs}|^2] = e.
\end{gather}

The condition (\ref{Eqn21}) requires that the functions $U_{qs}$
and $ V_{qs}$ should depend on the modulus of the vector $\vec{q}$
only. Besides, one should take into account that the trial vector
$|\Phi^{(0)}_1\rangle$ is not the accurate eigenvector of the
exact integrals of motion $\hat Q$ and $\hat{ \vec {P}}$ as it
represents the accurate eigenvector of the Hamiltonian
$|\Phi_1\rangle$ only approximately. Therefore, in the considered
zero approximation the conservation laws for momentum and charge
can be satisf\/ied only on  average, and this leads to the above
written normalized conditions. Generally, equation~(\ref{Eqn21a})
should not be considered as the additional condition for the
variational parameters but as the def\/inition of the observed
charge of the ``physical'' particle at the given value of the
initial charge of the ``bare'' particle. Therefore the sign of the
observed charge is not f\/ixed a priori. Calculating the
sequential approximations to the exact state vector
$|\Phi_1\rangle$ (see Section~\ref{sec4}) should restore the
accurate integral of motion as well. An analogous problem appears
in the ``polaron'' theory when the momentum conservation law was
taken into account for the case of the strong coupling (for
example, \cite{Bogol,Gross}).

The trial vector $|\Phi_1\rangle$ is actually the collective
excitation of the system and in this respect the variational
approach dif\/fers greatly from the perturbation theory where the
zero approximation for a one-particle state is described by one of
the following state vectors:
\begin{eqnarray}
\label{Eqn21b}  |\Phi^{(PT)}_1e\rangle = a^+_{\vec{P} s} | 0; 0; 0
\rangle, \qquad |\Phi_1\rangle \simeq |\Phi^{(PT)}_1p\rangle =
b^+_{\vec{P} s} | 0; 0; 0 \rangle.
\end{eqnarray}

These vectors do not depend on any parameters and are eigenvectors
of the momentum and charge operators. But they correspond to
one-particle excitations determined by the charge $e_0$ of the
``bare'' electron and the f\/ield $\varphi (\vec r ) = 0$. We
suppose that the introduction of the variational parameters into
the wave function of the zero approximation will enable us to take
into account the vacuum polarization.

It should be noticed that the another reason of inconsistency of
the states (\ref{Eqn21b}) as the ``physi\-cal'' electron states
because of the low frequency photon f\/ield was considered
recently in~\cite{physstate}.

So, the following variational estimation for the energy $E_1(0) =
E_1$ $(\vec{P}=0)$ of the state corresponding to the ``physical''
quasi-particle excitation of the whole system is considered in the
strong coupling zero approximation:
\begin{gather}
\label{Eqn22} E_1 (0) \simeq  E^{(0)}_1 [U_{qs};
V_{qs};\varphi(\vec{r})] = \langle \Phi^{(0)}_1|\hat H
|\Phi^{(0)}_1\rangle,
\end{gather}
where the average is calculated with the full
Hamiltonian~(\ref{Eqn18}) and the functions $U_{qs}$, $V_{qs}$,
$\varphi(\vec{r})$ are to be found as the solutions of variational
equations
\begin{gather}
\label{Eqn23} \frac{\partial E^{(0)}_1 (U_{qs};
V_{qs};\varphi(\vec{r}))}{\partial U_{qs}} = \frac{\partial
E^{(0)}_1 }{\partial V_{qs}} =  \frac{\partial E^{(0)}_1
}{\partial \varphi(\vec{r})} = 0
\end{gather}
with the additional conditions   (\ref{Eqn21}), (\ref{Eqn21a}).

It is quite natural, that the ground state energy is calculated
in the framework of the considered approximation as follows:
\begin{gather*}
\label{Eqn24} E_0 \simeq  E^{(0)}_0 = \langle\Phi_0|\hat H
|\Phi_0\rangle = 0.
\end{gather*}

Further discussion is needed in connection with the application of
the variational princip\-le~(\ref{Eqn22}),~(\ref{Eqn23}) for
estimating the energy of the excited state, because usually the
variational principle is used for estimating the ground state
energy only. As far as we know it was f\/irst applied
in~\cite{Caswell}  for nonperturbative calculation of the excited
states of the anharmonic oscillator with an arbitrary value of
anharmonicity. This approach was called the ``principle of the
minimal sensitivity''. It was shown in our paper~\cite{OM82}, that
the application of the variational principle to the excited states
is actually the consequence of the fact that the set of
eigenvalues for the full Hamiltonian does not depend on the choice
of the representation for eigenfunctions. As a result, the
operator method for solving Schr\"odinger equation was developed
as the regular procedure for calculating further corrections to
the zero-order approximation. Later this method was applied to a
number of real physical systems and proved to be very ef\/fective
when calculating the energy spectrum in the wide range of the
Hamiltonian parameters and quantum numbers (\cite{OM95,OM96,Acta}
and the cited references).

The average value in equation~(\ref{Eqn22}) is calculated
neglecting the classical components of the vector f\/ield. They
appear in the high-order corrections that  are def\/ined by the
renormalized charge $e \ll 1$ and can be considered by means of
the canonical perturbation theory (Section~\ref{sec5}). It means
that
\begin{gather*}
 \langle \Phi^{(0)}_1 | \hat \psi^* (\vec{r}) [ \vec \alpha \hat{ \vec{ A} } (\vec{r})] \hat \psi
(\vec{r})|\Phi^{(0)}_1\rangle  = 0.
\end{gather*}

It should be noted that the possibility of constructing
self-consistently the renormalized QED at the non-zero vacuum
value of the scalar f\/ield operator was considered before
\cite{Fradkin} but the solution  of the corresponding equations
was not  discussed.

Then the functional for determining the zero approximation for the
energy of the one-particle excitation is def\/ined as
follows:{\samepage
\begin{gather}
 E_1(0) = \int d \vec r \int \frac{d \vec q}{(2\pi)^{3/2}} \int \frac{d \vec {q}\,{}'}{(2\pi)^{3/2}}
\sum_{s,s'} \sum_{\mu,\nu} \{ U^*_{q's'} u^*_{\vec{q}\,{}' s' \mu}
[(\vec \alpha \vec q + \beta m_0)_{\mu \nu} + e_0 \varphi (\vec r)
\delta_{\mu \nu}]
U_{q s} u_{\vec{q} s \nu}  \nonumber\\
\phantom{E_1(0) =}{} -V_{q's'} v^*_{\vec{q}\,{}' s' \mu} [(\vec
\alpha \vec q + \beta m_0)_{\mu \nu} + e_0 \varphi (\vec r)
\delta_{\mu \nu}] V^*_{qs} v_{\vec{q} s \nu}\} e^{i (\vec q -
\vec{q}\,{}') \vec r} - \frac{1}{2} \int d \vec r [\vec{\nabla}
\varphi (\vec r)]^2.\label{Eqn25a}
\end{gather}}

In order to vary the introduced functional let us def\/ine the
spinor wave functions (not opera\-tors) which describe the
coordinate representation for the electron and positron wave
packets in the state vector  $ |\Phi^{(0)}_1\rangle$:
\begin{gather}
 \Psi_{\nu} (\vec r) = \int \frac{d \vec q}{(2\pi)^{3/2}} \sum_{s} U_{q s} u_{\vec{q} s \nu}
e^{i \vec q \vec r}, \qquad \Psi^c_{\nu} (\vec r) = \int \frac{d
\vec q}{(2\pi)^{3/2}} \sum_{s} V^*_{q s} v_{\vec{q} s \nu} e^{i
\vec q \vec r}.\label{Eqn26}
\end{gather}

In particular, if the trial state vector is chosen in one of the
forms (\ref{Eqn21b}) of the standard perturbation theory, the wave
functions  (\ref{Eqn26}) coincide with the plane wave solutions of
the free Dirac equation. For a general case the variation of the
functional (\ref{Eqn25a}) by the scalar f\/ield leads~to
\begin{gather}
 E(0) = \int d \vec r \{ \Psi^+ (\vec r) \left[(-i\vec \alpha \vec \nabla + \beta m_0) +
\frac{1}{2}e_0 \varphi (\vec r) \right] \Psi (\vec r)  \nonumber\\
\phantom{E(0) =}{} -\Psi^{+c} (\vec r) \left[(-i\vec \alpha \vec
\nabla + \beta m_0) +
\frac{1}{2}e_0 \varphi (\vec r) \right] \Psi^c (\vec r), \nonumber\\
\int {d \vec{r}}\, [\Psi^{+} (\vec r) \Psi (\vec r) + \Psi^{+c}
(\vec r\,{}') \Psi^{c} (\vec r\,{}')] = 1,\label{Eqn27}
\\
\label{Eqn27b} \varphi (\vec r) = \frac{e_0}{4 \pi} \int \frac {d
\vec{r}\,{}'} {|\vec r - \vec{r}\,{}'|} [\Psi^{+} (\vec r\,{}') \Psi
(\vec r\,{}') - \Psi^{+c} (\vec r\,{}') \Psi^{c} (\vec r\,{}')].
\end{gather}

The main condition for the existence of the considered
nonperturbative excitation in QED is def\/ined by the extremum of
the functional (\ref{Eqn27}) corresponding to a non-zero classical
f\/ield. The structure of this functional shows that such
solutions of the variational equations could appear only if the
trial state vector simultaneously included the superposition of
the electron and positron wave packets. So, such solutions cannot
be obtained by means of the perturbation theory with the state
vectors (\ref{Eqn21b}).

Equation (\ref{Eqn27}) and the Fourier representation
(\ref{Eqn20}) for the trial vector clearly indicate that the
assumption concerning the localization of the functions $\Psi
(\vec r)$ near some point does not contradict to the translational
symmetry of the system because this point by itself  can   be
situated at any point of the full space with equal probability.
The general analysis of the correlation between the local breaking
of the symmetry and the conservation of accurate integral of
motion for the arbitrary quantum system was considered in detail
by Bogoluibov in his widely known paper ``On
quasi-averages''~\cite{quasi}. A similar analysis of the problem
in question will be given in Section~\ref{sec4}.

Varying the functional (\ref{Eqn27}) by the wave functions $\Psi
(\vec r)$ and $\Psi^c (\vec r)$ taking into account their
normalization conditions leads to the equivalent Dirac equations
describing the electron (positron) motion in the f\/ield of
potential $\varphi (\vec r)$:
\begin{gather}
 \{(-i\vec \alpha \vec \nabla + \beta m_0) +
e_0 \varphi (\vec r)  \} \Psi (\vec r) = 0,\qquad \{(-i\vec \alpha
\vec \nabla + \beta m_0) + e_0 \varphi (\vec r)  \} \Psi^{c} (\vec
r) = 0.\label{Eqn27a}
\end{gather}

But it is important that in spite of the normalization condition
(\ref{Eqn27}) for the total state vector~(\ref{Eqn26}) each of its
components could be normalized dif\/ferently
\begin{gather}
 \int {d \vec{r}} \Psi^{+} (\vec r) \Psi (\vec r) = \frac{1}{1 + C},\qquad
\int {d \vec{r}}\Psi^{+c} (\vec r\,{}') \Psi^{c} (\vec r\,{}') =
\frac{C}{1 + C}.\label{Eqn27d}
\end{gather}

The constant  $C$ is an arbitrary value up to now. It def\/ines
the ratio of two charge states in the considered wave packet. As a
result the self-consistent potential  $\varphi (\vec r)$ of the
scalar f\/ield depends on  $C$ because of the
equation~(\ref{Eqn27b}).

We should discuss the procedure of separating variables in more
detail, because of the non-linearity of the obtained system of
equations for the wave functions and the self-consistent
potential. Since the considered physical system has no preferred
vectors if  $\vec P = 0$, it is natural to regard the
self-consistent potential as spherically symmetrical. Then the
variable separation for the Dirac equation is realized on the
basis of the well known spherical bispinors \cite{Akhiezer}:
\begin{gather*}
\Psi_{jlM} = \left( \begin{array}{c}
g(r) \Omega_{jlM}\\
i f(r) \Omega_{jl'M}
\end{array} \right).
\end{gather*}

Here $\Omega_{jlM}$ are the spherical spinors \cite{Akhiezer}
describing the spin and angular dependence of the one-particle
excitation wave functions; $j$, $M$ are the total excitation
momentum and its projection respectively, the orbital momentum
eigenvalues are connected by the correlation $ l + l' = 2j$. It is
natural to consider the state with the minimal energy as the most
symmetrical one, corresponding to the values  $j=1/2$, $M = \pm
1/2$, $l = 0$; $l' = 1$. This choice corresponds to the condition
according to which in the non-relativistic limit the ``large''
component of the bispinor $\Psi$ $\sim g$ corresponds to the
electronic zone of the electron-positron f\/ield. Then the unknown
radial functions $f$, $g$ satisfy the following system of the
equations:
\begin{gather}
 \frac{d (rg)}{dr} - \frac{1}{r}(rg) - (  m_0 - e_0 \varphi(r)) (rf) = 0,
\nonumber\\
\frac{d (rf)}{dr} + \frac{1}{r}(rf) - (  m_0 + e_0 \varphi(r))
(rg) = 0.\label{Eqn29}
\end{gather}

The states with various projections of the total momentum should
be equally populated in order to be consistent with the assumption
of the potential spherical symmetry with the
equation~(\ref{Eqn27b}). So, the total wave function of the
``electronic'' component of the quasi-particle excitation of the
electron positron f\/ield is chosen in the following form:
\begin{gather}
 \Psi = \frac{1}{\sqrt{2}} [\Psi_{1/2,0,1/2} + \Psi_{1/2,0,-1/2}]= \left(
\begin{array}{c}
g(r) \chi^+_0\\
i f(r) \chi^+_1
\end{array} \right),
\nonumber\\
\chi^+_l =\frac{1}{\sqrt{2}} [\Omega_{1/2,l,1/2} +
\Omega_{1/2,l,-1/2}], \qquad l = 0,1.\label{Eqn28a}
\end{gather}

In its turn, the wave function  $\Psi^c$ is def\/ined on the basis
of the following bispinor:
\begin{gather}
\label{Eqn30} \Psi^c_{jlM} = \left( \begin{array}{c}
- i f_1(r) \Omega_{jlM}\\
g_1(r) \Omega_{jl'M}
\end{array} \right).
\end{gather}

The radial wave functions  $f_1$, $g_1$ in this case satisfy the
following system of equations
\begin{gather}
 \frac{d (rg_1)}{dr} + \frac{1}{r}(rg_1) - (  m_0 + e_0 \varphi(r)) (rf_1) = 0,
\nonumber\\
\frac{d (rf_1)}{dr} - \frac{1}{r}(rf_1) - (  m_0 - e_0 \varphi(r))
(r g_1) = 0.\label{Eqn31}
\end{gather}

These equations correspond to  the positronic branch of the
electron-positron f\/ield with the ``large'' component $\sim g_1$
in the non-relativistic limit.

It is important to note that the functions  $\Psi$ and $\Psi^c$
satisfy the equations (\ref{Eqn27a}) that have the same form. It
imposes an additional condition of the orthogonality on them:
\begin{gather}
\label{Eqn31a} \langle\Psi^c|\Psi\rangle  = 0.
\end{gather}

Taking into account this condition and also the requirement that
the states with  dif\/ferent values of $M$ should be equally
populated one f\/inds the ``positronic'' wave function{\samepage
\begin{gather}
 \Psi^c = \frac{1}{\sqrt{2}} [\Psi^c_{1/2,0,1/2} - \Psi^c_{1/2,0,-1/2}]= \left(
\begin{array}{c}
-i f_1(r) \chi^-_0\\
g_1(r) \chi^-_1
\end{array} \right ),
\nonumber\\
\chi^-_l =\frac{1}{\sqrt{2}} [\Omega_{1/2,l,1/2} -
\Omega_{1/2,l,-1/2}], \qquad l = 0,1.\label{Eqn31b}
\end{gather}}

The equation for the self-consistent potential follows from the
def\/inition of $\varphi (r)$ in  formula~(\ref{Eqn27b}) taking
into account the normalization of the spherical spinors
\cite{Akhiezer}:
\begin{gather}
\label{Eqn32} \frac{d^2 \varphi}{d r^2} + \frac{2}{r}\frac{d
\varphi}{d r} = - \frac{e_0}{4\pi}[f^2 + g^2 - f_1^2 - g_1^2].
\end{gather}

The boundary condition for the potential is equivalent to the
normalization condition (\ref{Eqn21a}) and def\/ines the charge
$e$ of the ``physical'' electron (positron)
\begin{gather}
\label{Eqn33} \varphi (r)|_{r \rightarrow \infty} = \frac{e}{4 \pi
r} = \frac{e_0}{4\pi r}\int_{0}^{\infty} r^2_1 dr_1[f^2(r_1) +
g^2(r_1) - f_1^2(r_1) - g_1^2(r_1)].
\end{gather}

It is important to stress that the form of the functions given
above is def\/ined practically uniquely by the imposed conditions.
At the same time the obtained equations are consistent with the
symmetries def\/ined by the physical properties of the system. The
f\/irst symmetry is quite evident and relates to the fact that the
excitation energy does not depend on the choice of the
quantization axis of the total angular momentum.

Moreover, these equations satisfy the condition of the charge
symmetry \cite{Akhiezer}. Indeed, by direct substitution, one can
check  that one more pair of bispinors leads to the equations
completely coinciding with (\ref{Eqn29}), (\ref{Eqn31})
\begin{gather}
\label{Eqn33a}\tilde{ \Psi}_{jlM} = \left( \begin{array}{c}
i g_1(r) \Omega_{jl'M}\\
- f_1(r) \Omega_{jlM}
\end{array} \right),
\\
\label{Eqn33b}\tilde{ \Psi}^c_{jlM}= \left( \begin{array}{c}
- f(r) \Omega_{jl'M}\\
- i g(r) \Omega_{jlM}
\end{array} \right).
\end{gather}

It means that these bispinors allow one to f\/ind another pair of
the wave functions which are orthogonal to each other and to the
functions  (\ref{Eqn28a}), (\ref{Eqn31b}) but include the same set
of the radial functions
\begin{gather}
\label{Eqn33c} \tilde{\Psi} = \left(
\begin{array}{c}
i g_1(r) \chi^-_1\\
-f_1(r) \chi^-_0
\end{array} \right),\qquad
\tilde{\Psi}^c =\left(
\begin{array}{c}
- f(r) \chi^+_1\\
-i g(r) \chi^+_0
\end{array} \right).
\end{gather}

These functions dif\/fer from the set (\ref{Eqn28a}),
(\ref{Eqn31b}) because they lead to a dif\/ferent sign of the
observed charge of the quasi-particle due to the boundary
condition  (\ref{Eqn33}) and describe the ``physi\-cal'' positron.

The structure of the equation   (\ref{Eqn27a}) shows that the
considered variational method is consistent with the gauge
symmetry of the initial Hamiltonian. One can see that these
equations are invariant with respect to the following
transformations:
\begin{gather*}
   \vec \nabla \Rightarrow \vec \nabla + i e_0 \vec A_l (\vec r) ,\nonumber\\
 \Psi (\vec r) \Rightarrow e^{- i \beta (\vec r)}  \Psi (\vec r), \qquad \vec \nabla \beta (\vec r) = e_0 \vec A_l (\vec  r),
\end{gather*}
 with an arbitrary longitudinal potential   $\vec A_l (\vec r)$.

It means that the Hamiltonian   (\ref{Eqn18}) could be chosen in
an arbitrary Lorentz gauge with the classical components both for
the scalar f\/ield   $\varphi (\vec r)$ and for the longitudinal
f\/ield $\vec A_l (\vec r)$ if the following condition was
fulf\/illed:
\begin{gather*}
   \Delta \beta (\vec r) = \rho_l (\vec r) = 0,\qquad \rho_l (\vec r) =  \Psi^+ (\vec r) (\vec r \cdot \vec \nabla)  \Psi (\vec r).
\end{gather*}

One can easily check that the condition   $\rho_l (\vec r) = 0$ is
fulf\/illed identically for the functions that describes the
quasi-particle above.

Let us now proceed to the solution of the variational equations.
It follows from the  qualitative analysis that the important
property of the trial state vector is the possibility to vary the
relative contribution of the electronic and positronic components
of the wave function. Therefore let us introduce the variational
parameter $C$ in the following way:
\begin{gather*}
\int_{0}^{\infty} r^2 dr[f^2(r) + g^2(r)] = \frac{1}{1 + C},
\qquad \int_{0}^{\infty} r^2 dr[f_1^2(r) + g_1^2(r)] = \frac{C}{1
+ C}.
\end{gather*}

The dimensionless variables and new functions can be introduced
\begin{gather}
x = r m_0, \qquad E = \epsilon m_0, \qquad  e_0 \varphi(r) =
m_0\phi(x),
\qquad \frac{e^2_0}{4 \pi} = \alpha_0, \qquad u(x) \sqrt{m_0} = r g(r),\nonumber\\
v(x)\sqrt{m_0} = r f(r),\qquad u_1(x)\sqrt{m_0} = r g_1(r),\qquad
v_1(x)\sqrt{m_0} = r f_1(r).\label{Eqn35}
\end{gather}

As a result the system of equations for describing the radial wave
functions of the   one-particle excitation of the
electron-positron f\/ield and the self-consistent potential of the
vacuum polarization can be obtained:
\begin{gather}
 \frac{d u}{dx} - \frac{1}{x}u - (  1 - \phi(x)) v  = 0,
\qquad \frac{d v}{dx} + \frac{1}{x}v - (  1 + \phi(x)) u = 0,
\nonumber\\
\frac{d u_1}{dx} + \frac{1}{x}u_1 - (  1 + \phi(x)) v_1 = 0,
\qquad \frac{d v_1}{dx} - \frac{1}{x}v_1 - (\  1 - \phi(x)) u_1 =
0,\label{Eqn36}
\\
\phi(x) = \alpha_0 \left[ \int_{x}^{\infty} dy \frac{\rho(y)}{y} +
\frac{1}{x} \int_{0}^{x} dy \rho(y)\right],\qquad \rho(x) =
\left[u^2(x) + v^2(x) - u^2_1(x) - v^2_1(x)\right].\nonumber
\end{gather}

The mathematical structure of equations (\ref{Eqn36}) is analogous
to that of the self-consistent equations for localized state of
``polaron'' in the strong coupling limit~\cite{Bogol}. Therefore
the same approach can be used for the numerical solution of these
nonlinear equations. It has been developed and applied
\cite{Krylov} for the ``polaron'' problem on the basis of the
continuous analog of Newton's method.

Let us take into account that the system of equations
(\ref{Eqn36}) can be simplif\/ied because the pairs of the
functions $u$, $v$ and $u_1$, $v_1$ are satisf\/ied by the same
equations and dif\/fer only by the normalized condition. Therefore
they can be represented by means of one pair of functions if
special notations are used:
\begin{gather}
 u = \sqrt{\frac{1}{1 + C}} u_0,\qquad  v = \sqrt{\frac{1}{1 + C}} v_0 ,
\qquad u_1 = \sqrt{\frac{C}{1 + C}} v_0, \qquad  v_1 =
\sqrt{\frac{C}{1 + C}} u_0 ,
\nonumber\\
\int_{0}^{\infty}  dx [u_0^2(x) + v_0^2(x)] = 1,\qquad  \rho_0(x)
= u_0^2(x) + v_0^2(x),
\nonumber\\
\frac{d u_0}{dx} - \frac{1}{x}u_0 - ( 1 - \phi(x)) v_0  = 0,
\qquad
\frac{d v_0}{dx} + \frac{1}{x}v_0 - ( 1 + \phi(x)) u_0 = 0, \nonumber\\
\phi(x) = \alpha_0 \frac{1 - C}{1 + C}\phi_0(x),\qquad \phi_0(x) =
\int_{x}^{\infty} dy \frac{\rho_0(y)}{y} + \frac{1}{x}
\int_{0}^{x} dy \rho_0(y).\label{36}
\end{gather}

The energy of the system (\ref{Eqn27}) can also be calculated by
these functions:
\begin{gather}
  E_1(0)\equiv E(0) = m_0 \frac{1 - C}{1 + C}\left[ T + \frac{1}{2}\alpha_0 \frac{1 - C}{1 + C} \Pi\right], \nonumber\\
T = \int_{0}^{\infty} dx [ (u_0' v_0 - v_0' u_0) - 2 \frac{u_0
v_0}{x} + (u_0^2 - v_0^2)], \qquad \Pi = \int_{0}^{\infty} dx
\phi_0 (u_0^2 + v_0^2).\label{37}
\end{gather}
and equation~(\ref{36}) can be obtained when varying of the
functional~(\ref{37}).

The required solutions are to be normalized and this condition
def\/ines the asymptotic behavior of the functions near the
integration interval boundaries:{\samepage
\begin{gather*}
u_0 \approx A x\left[ 1 + \frac{1 - \phi^2(0)}{6}x^2\right] ,
\qquad  v_0 \approx A \frac{1 - \phi(0)}{3} x^2, \qquad x
\rightarrow 0,
\nonumber\\
u_0 \approx  A_1 \exp(-x), \qquad v_0 \approx  - A_1
\exp(-x),\qquad x \rightarrow \infty.
\end{gather*}}

The value
\begin{gather}
\label{38}  a = \alpha_0 \frac{1 - C}{1 + C},
\end{gather}
is the free parameter of the equations (\ref{36}) and it plays a
role of the eigenvalue when the nontrivial normalized solution
exists.

The method for the numerical solution of the nonlinear
self-consistent system of the equations~(\ref{36}) was described
in detail in the paper \cite{SLS}. Only the numerical results for
the localized wave functions and for the scalar potential are
described in the present work. Certainly, the numerical value for
the parameter  $a$ depends on the accuracy of the
f\/inite-dif\/ference approximation for the dif\/ferential
operators in the whole interval of integration. The value $a$ was
calculated more accurately in comparison with~\cite{SLS}:
\begin{gather}
\label{38a}  a = a_0 \approx - 3.531.
\end{gather}

\begin{figure}[t]
\begin{minipage}[t]{75mm}\centering
\includegraphics [width=6.8cm]{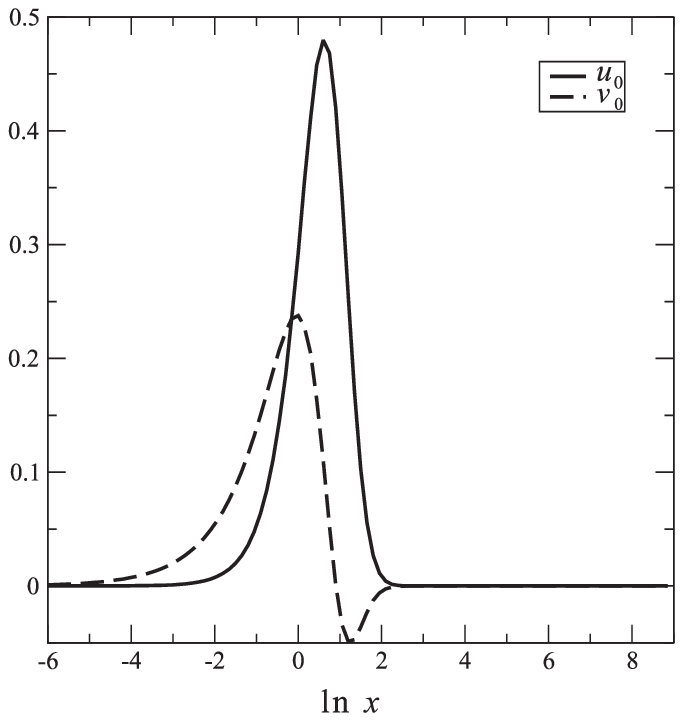}
\caption {Localized wave functions of the quasi-particle
excitation; $x = r/r_0= \frac{2|a_0|r}{ r_e}$, $u_0 = \xi  x g(x
r_0)$, $v_0 = u_0 = \xi  x f(x r_0)$, $\xi =
\frac{1}{4}(\frac{\alpha}{|a_0| m })^{3/2}$.} \label{Fig.1}
\end{minipage}\hfill
\begin{minipage}[t]{75mm}\centering
\includegraphics[width=7.45cm]{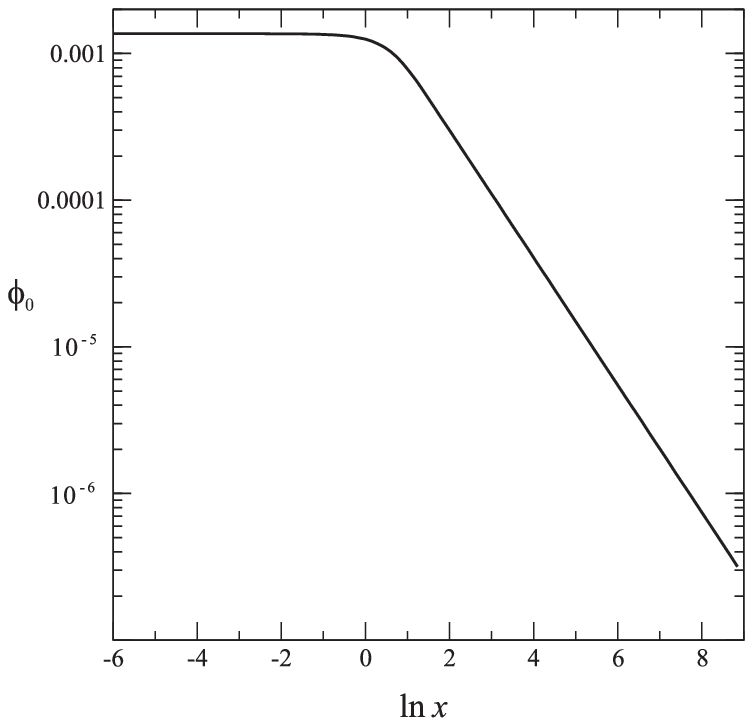}  \caption {Self-consistent potential of the excitation;
$x = r/r_0 = \frac{2|a_0|r}{ r_e}$,  $\phi_0 (x) = \frac{e}{2
|a_0| m }\varphi(x r_0)$.} \label{Fig.2}
\end{minipage}
\end{figure}

Fig.~\ref{Fig.1} shows the solutions $u_0$, $v_0$ for the electron
and positron components of the excitation that are localized in
the domain with the linear size of $\sim  m_0^{-1}$.
Fig.~\ref{Fig.2} represents the self-consistent potential $\phi_0$
that provides   stability of the system in this domain and
corresponds to the value $a_0$. It gets over the Coulomb potential
of the ``physical'' charge $e$ for $r > r_0 = m_0^{-1}$. It is
important that the characteristic size of this excitation $r_0$ is
the same order as the classical radius of the electron $r_e  =
\alpha/m$, namely $r_0 = \frac{r_e}{2|a_0|}\approx 0.15 r_e$  (see
below equation~(\ref{44})).

\section[Physical interpretation of the quasi-particle excitation and estimation of the $\mu$-meson mass]{Physical interpretation of the quasi-particle excitation\\ and estimation of the $\boldsymbol{\mu}$-meson mass}\label{sec3}

The stationary localized collective excitation of the
electron-positron f\/ield described above is of great interest by
itself as the eigenvector of the well known QED Hamiltonian that
cannot be calculated by means of the perturbation theory and has
not be considered before. But it is also essential to f\/ind its
physical interpretation because the only   stable objects observed
in the electrodynamic processes are electrons (positrons) and
photons. Therefore it is natural to suppose that this localized
state describes the ``physical'' electron (positron) with the
observed charge~$e$. The integral charge of the considered
one-particle excitation is def\/ined by the boundary condition
(\ref{Eqn33}) and this supposition leads to:
\begin{gather}
\label{39} e_0 \frac{(1-C)}{(1+C)} = e.
\end{gather}

Taking into account the def\/inition (\ref{38}) one can found the
following relation between the ``primary'' coupling constant
$\alpha_0 = e_0^2/4\pi$ and the observed value of the f\/ine
structure constant   $\alpha = e^2/4\pi$
\begin{gather}
\label{40} \alpha_0 = \frac{a_0^2}{\alpha} \approx 1708.1.
\end{gather}

This formula def\/ines the renormalization of the charge in the
considered approximation and shows self-consistency of the initial
supposition that the interaction between the ``primary''
electron-positron and scalar f\/ields   is strong. Then the
renormalization constant \cite{Akhiezer} ($\alpha = Z(\alpha)
\alpha_0$)~is:
\begin{gather}
\label{40a} Z^{(0)}(\alpha) = \frac{\alpha^2 }{a_0^2}.
\end{gather}

It should be stressed once more that the large value of the
``primary'' coupling constant $\alpha_0$ does not mean at all that
the perturbation theory cannot be applied for the calculation of
the observed physical values. In connection with it let us remind
that the ``primary'' coupling constant in the existed form of QED
tends to inf\/inity ($\alpha_0 \rightarrow \infty$) because of
``Landau pole'' \cite[\S~128]{Lifshitz}. Nevertheless, it can be
excluded from the observed values by means of the renormalization
procedure. In our representation the large but f\/inite value of
$\alpha_0$ is important only for the formation of the initial
basis of the self-localized states but it is also excluded when
calculating the observed physical values in a power series of the
``physical'' coupling constant $\alpha \simeq \alpha_0^{-1} \ll
1$. However, in this case one can avoid the divergent integrals
when performing the renormalization procedure (see below
Section~\ref{sec5}).

The QED Hamiltonian (\ref{Eqn18}) is def\/ined with  $e_0> 0$ and
the integral charge of the quasi-particle $e < 0$ because of the
conditions  (\ref{38}) and  (\ref{38a}). It means that the
considered excitation corresponds to the ``physical'' electron.
The excitation corresponding to the ``physical'' positron is
def\/ined by the bispinors (\ref{Eqn33a}).

The change of the integral charge of the excitation is explained
by the local intersection of the electron and positron energy
zones in the strong self-consistent scalar electromagnetic
f\/ield. The analogous states appear also in the strong Coulomb
f\/ield of the nucleus with the large charge  $Z > 137 $ when the
gap between Dirac zones tends to zero and stabilization of the
electron states is achieved due to creation of the additional
positrons (see, for example, \cite{critical} and references
therein). In our case the origin of the charge density of one sign
is the source of the strong scalar electromagnetic f\/ield that
leads to coming together the electron  and positron levels. This
f\/ield can be compensated by the extra value of the charge
density of opposite sign. More detailed qualitative analysis of
the structure of the quasi-particle excitation  was discussed in
the paper \cite{renorm}. It is also interesting to note that the
relation between the charges of the ``bare'' and ``physical''
electrons has the same form as the equation for the electric and
magnetic charges in the theory of the Dirac monopole
\cite{monopol}.

The eigenvalue $a_0$ corresponds to minimum of the functional
(\ref{37}) with f\/ixed functions   $u_0$, $v_0$, $\varphi_0$ with
respect to the parameter  $\xi = (1-C)/(1+C)$ and can be expressed
in terms of the integrals $T$ and $\Pi$ that def\/ine  kinetic and
potential contributions in the total excitation energy
\begin{gather}
\label{41} a_0 = - \frac{T}{\Pi},
\end{gather}
and the numerical value of the integral  $T$ can be calculated
with the considered accuracy as
\begin{gather}
\label{42} T \approx 0.749.
\end{gather}

Then the total energy of the excitation with zero momentum is:
\begin{gather}
\label{43} E(  0) = - \frac{m_0}{\alpha_0} \frac{T a_0}{2} = - m_0
\alpha \frac{T}{2 a_0} > 0.
\end{gather}

This value def\/ines the minimal energy of the one-particle
excitation of the electron-positron f\/ield and its positive sign
corresponds to the ``bottom'' of the ``physical'' electron zone in
the renormalized QED. It is also consistent with the negative
charge of this excitation  \cite{renorm}. It is natural to
consider this value as the rest-mass   $m$ of the ``physical''
electron and to f\/ind the mass renormalization in QED:
\begin{gather}
 E(  0) = m \equiv m_e   = - m_0 \alpha \frac{T}{2 a_0},\qquad
m_0 = m_e \frac{2 |a_0|}{\alpha} \approx 1291.7 m_e.\label{44}
\end{gather}

This relation ($E(  0) \ll m_0$ ) shows that the primary mass of
the ``bare'' electrons and positron~$ \sim m_0$ is compensated by
the binding energy of their charge distributions almost
completely.  It is interesting that the characteristic size of the
domain, where the considered one-particle excitation is localized
($\Delta r \approx m_0^{-1}$), is the same order as the value of
the electron ``radius'' ($r_e = \alpha/m$) in the classical model
of Abraham--Lorentz~\cite{Dirac1}.

In accordance with the formulas (\ref{40}), (\ref{44})
characteristics of the charge spatial distribution   in the
``physical'' electron depend on the observable QED parameters
$\alpha$ and $m$ only. Such internal structure of the electron
does not contradict to the well known results (for
example,~\cite{Brodsky}) that maintain that the observed
characteristics of the electron cannot depend on any dimensional
parameter with the exception of $m$ (see below the discussion in
Section~\ref{sec5}). Actually the local charge distribution in QED
arises as well in the framework of the perturbation theory   when
considering the physical interpretation of the renormalization of
charge (for example,~\cite{Akhiezer}).

As it was shown by Dirac  \cite{Dirac1}, investigation of the
``physical'' electron with the distributed charge is of great
interest because it gives the possibility to interpret  the
``physical''   $\mu$-meson as the excited state of such system.
In order to describe the dynamics of such excitation Dirac
introduced the hypothetical elastic parameter that is absent in
QED. However, the variational approach considered in the present
paper allows one to analyze   the one-particle excitation
dif\/fered from the ``physical'' electron without inclusion of any
additional parameters.

Let us choose the trial state vector in the same form as it is
given by formulas   (\ref{Eqn20}), (\ref{Eqn22}), but with the
wave functions that are orthogonal to the ``physical'' electron
state vector. These functions are satisf\/ied to the system of the
self-consistent equations in the form
(\ref{Eqn27})--(\ref{Eqn27d}), but with the nonzero eigenvalues
$E_1 \neq E_2$ for both components of the ``primary''
electron-positron f\/ield:
\begin{gather}
 \{(-i\vec \alpha \vec \nabla + \beta m_0) +
e_0 \varphi_{\mu} (\vec r) - E_1 \} \Psi_{\mu} (\vec r) = 0,\nonumber\\
\{(-i\vec \alpha \vec \nabla + \beta m_0) + e_0 \varphi_{\mu}
(\vec r) - E_2 \} \Psi^{c}_{\mu} (\vec r) = 0,
\nonumber\\
\varphi_{\mu} (\vec r) = \frac{e_0}{4 \pi} \int \frac {d \vec{r}\,{}'}
{|\vec r - \vec{r}\,{}'|}
[\Psi^{+}_{\mu} (\vec r\,{}') \Psi_{\mu} (\vec r\,{}') - \Psi^{+c}_{\mu} (\vec r\,{}') \Psi^{c}_{\mu} (\vec r\,{}')],\nonumber\\
\int {d \vec{r}} \, [\Psi^{+}_{\mu} (\vec r) \Psi_{\mu} (\vec r) +
\Psi^{+c}_{\mu} (\vec r\,{}') \Psi^{c}_{\mu} (\vec r\,{}')] = 1,
\qquad \int {d \vec{r}}\, [\Psi^{+}_{\mu} (\vec r) \Psi  (\vec r) +
\Psi^{+c}_{\mu} (\vec r\,{}') \Psi^{c}
(\vec r\,{}')] = 0, \nonumber\\
\int {d \vec{r}} \Psi^{+}_{\mu} (\vec r) \Psi_{\mu} (\vec r) =
\frac{1}{1 + C}, \qquad \int {d \vec{r}}\Psi^{+c}_{\mu} (\vec
r\,{}') \Psi^{c}_{\mu} (\vec r\,{}') = \frac{C}{1 + C}.\label{45}
\end{gather}

In this case the energy that def\/ines the observed mass of the
``physical''   $\mu$-meson can be calculated as
\begin{gather*}
E_{\mu}(\vec P = 0) = m_{\mu}   = E_1 - E_2.
\end{gather*}

The parameters $C$, $e_0$, $m_0$ in equations (\ref{44}) should be
the same as was def\/ined by equations~(\ref{39})--(\ref{44})
because the observed charges of the ``physical'' electron and
$\mu$-meson coincide. Besides, the same bispinors
(\ref{Eqn28a})--(\ref{Eqn33b}) as for electron should be used when
separating variables in equation~(\ref{45}). However, the radial
functions for both components of the ``primary'' electron-positron
f\/ield that form the ``physical'' $\mu$-meson should correspond
to the dif\/ferent eigenvalues.

If the dimensionless variables   (\ref{Eqn35}) have been used the
following system of equations is obtained:
\begin{gather}
\frac{d u_{\mu}}{dx} - \frac{1}{x}u_{\mu} - (  1 + \epsilon_1 -
\phi_{\mu}(x)) v_{\mu}  = 0, \qquad \frac{d v_{\mu}}{dx} +
\frac{1}{x}v_{\mu} - (  1 - \epsilon_1 + \phi_{\mu}(x)) u_{\mu} =
0,
\nonumber\\
\frac{d u_{1\mu}}{dx} + \frac{1}{x}u_{1\mu} - (  1 - \epsilon_2 +
\phi_{\mu}(x)) v_{1\mu} = 0, \qquad \frac{d v_{1\mu} }{dx} -
\frac{1}{x}v_{1\mu} - (  1 + \epsilon_2  - \phi_{\mu}(x)) u_{1\mu}
= 0,\!\!\!\label{47}
\\
\phi_{ \mu}(x) = \alpha_0 \left[ \int_{x}^{\infty}\! dy
\frac{\rho_{ \mu}(y)}{y} + \frac{1}{x} \int_{0}^{x} \!dy \rho_{
\mu}(y)\right]\!,\!\qquad\! \rho(x) = \left[u_{ \mu}^2(x) + v_{
\mu}^2(x) - u^2_{ 1\mu}(x) - v^2_{ 1\mu}(x)\right].\!\nonumber
\end{gather}

If the solution of the nonlinear problem (\ref{47}) with the
eigenvalues $\epsilon_{1,2}$ and the orthogonality and
normalization conditions is found, the observed mass of
$\mu$-meson is def\/ined by the following formula:
\begin{gather*}
   m_{\mu} = m_0 ( I_{\mu} - I_{1\mu}),\nonumber\\
I_{\mu} = \int_0^{\infty} dx \left\{ u'_{\mu} v_{\mu} -  u'_{\mu}
v_{\mu} - 2\frac{ u _{\mu} v_{\mu}}{x} + (u^2_{\mu}-
v^2_{\mu}) + \frac{1}{2}a_0\phi_{ \mu}(u^2_{\mu}+ v^2_{\mu})\right\},
\end{gather*}
where integral $I_{1\mu}$ is expressed by the functions $u
_{1\mu}$, $v_{1\mu}$ with the same formula as the
integral~$I_{\mu}$ by the functions $u _{ \mu}$, $v_{ \mu}$; the
parameter $a_0$ is def\/ined by equation~(\ref{41}).

Equation~(\ref{43}) for the observed mass of the electron can be
written in the same form
\begin{gather*}
 m_{e} = m_0 \frac{1 - C}{1 + C}  I_{0},
\end{gather*}
where integral $I_{0}$ is expressed by the functions $u _{0}$,
$v_{0}$ from equations (\ref{Eqn36}).

Calculation of the excited states for the nonlinear equations
(\ref{47}) with the self-consistent potential proved to be  a
quite dif\/f\/icult numerical problem. In the  present paper this
solution was calculated on the basis of the ``adiabatic''
approximation for the potential. It means that the potential was
f\/ixed in the same form as in equations~(\ref{Eqn36}): $\phi_{
\mu} \approx \phi_{0}$. In this case the functions $ u _{ \mu}$,
$v_{ \mu}$ coincide with $u _{0}$, $v_{0}$ but the functions $ u
_{ 1\mu}$, $v_{ 1\mu}$ correspond to the f\/irst excited state in
the Dirac equations   (\ref{Eqn36}). Then the relation (\ref{39})
and the def\/initions of the integrals   $I$ lead to the following
formulas:
\begin{gather*}
\frac{1 - C}{1 + C} = \frac{e}{e_0} \ll 1, \qquad  I_{0} \approx
I_{\mu},
\end{gather*}
and the observed electron-$\mu$-meson mass ratio can be calculated
as:
\begin{gather*}
\frac{m_{ \mu}}{m_{ e}} \approx  \frac{1 - C}{1 + C} \frac{I_{\mu}
- I_{1\mu}}{2 I_{\mu}}= \frac{|a_0|}{2\alpha}  \frac{I_{\mu} -
I_{1\mu}}{ I_{\mu}}\approx 242.2 \frac{I_{\mu} - I_{1\mu}}{
I_{\mu}}.
\end{gather*}

The calculated values for the integrals are the following
$I_{\mu}\approx 0.95$, $I_{1\mu}\approx 0.25$ and the mass ratio is
estimated as
\begin{gather*}
 \frac{m_{ \mu}}{m_{ e}} \approx  194.
\end{gather*}

This number can be compared with the experimental value   $(m_{
\mu}/m_{ e})_{E}\approx 206$ and the result obtained by Dirac
\cite{Dirac1}: $(m_{ \mu}/m_{ e})_{D}\approx 54$. Thus, the
physical interpretation of the QED one-particle excitations leads
to a quite good estimation for the observed mass of $\mu$-meson.
It can be improved if the completely self-consistent solution is
found for equations~(\ref{47}).

It may seem that interpretation of the ``physical'' $\mu$-meson as
the excited state of the ``bare'' electron-positron and
electromagnetic f\/ields contradicts to the experimentally
observed conservation of the muon (not electrical) charge in the
electromagnetic processes. However, let us remind that this  state
is the collective(!) excitation of the whole system with the
scalar electromagnetic f\/ield $\varphi_{\mu}$ not equal to the
f\/ield $\varphi$ for the ``physical'' electron self-localized
state. In accordance with the formula (\ref{Eqn20}) it means that
the both states are def\/ined by the coherent states corresponding
to dif\/ferent f\/ield amplitudes. Therefore the transition
between these states is def\/ined by the overlapping integral
between the corresponding coherent states of the electromagnetic
f\/ield. If one uses the standard def\/inition of such states
\cite{Akhiezer} this integral can be represented in the following
form
\begin{gather}
\label{52a}   I = \exp \left[ - 4 \pi \int \frac{d \vec
k}{\omega_k}|\vec E (\vec k) - \vec E_{\mu} (\vec k)|^2\right].
\end{gather}

Here $\vec E (\vec k)$, $\vec E_{\mu} (\vec k)$ are the Fourier
components of the corresponding electric f\/ield strengths, for
example:
\[
\vec E (\vec k) = - \frac{1}{(2\pi)^3} \int d \vec r \vec \nabla
\varphi (\vec r) e^{i \vec k \cdot \vec r}.
\]

With utilisation of  the dimensionless variables (\ref{Eqn35}) the
integral (\ref{52a}) can be written as
\begin{gather}
   I = e^{- \alpha_0 \Lambda},\qquad
\Lambda = \int_0^{\infty} t dt [\Phi(t) - \Phi_{\mu}(t)]^2, \nonumber\\
\Phi(t) = \int_0^{\infty} x dx \phi'(x) \sin (t x), \qquad
\Phi_{\mu}(t) = \int_0^{\infty} x dx \phi_{\mu}'(x) \sin (t
x).\label{52b}
\end{gather}

Parameter $ \Lambda \sim 1$ is def\/ined by the converged
integrals because of the asymptotic behavior of both functions
$\phi(x)$, $\phi_{\mu}(x)$  when $x \rightarrow \infty$ and $x
\rightarrow 0$. The overlapping integral (\ref{52b}) permits one
to estimate the electromagnetic lifetime of $\mu$-meson as
\begin{gather*}
\tau^{el}_{\mu} \sim m_0^{-1} I^{-1} \sim 10^{300} [{\rm sec}].
\end{gather*}

This value ensures the conservation of the muon charge in the
electromagnetic processes because it is essentially bigger than
the total $\mu$-meson lifetime conditioned by the weak
interaction.

\section{Lorentz invariance of the self-localized state\\ with nonzero momentum}\label{sec4}

In the previous sections the resting quasi-particle with a
non-trivial self-consistent charge distribution, the f\/inite
energy $E(0)$ and a zero total momentum $\vec P = 0$ was
considered in the framework of a nonperturbative QED.

The obtained solution allows one to imagine the internal structure
of the resting ``physical'' electron (positron) as a strongly
coupled state of charge distributions of the opposite sign. The
large values of integral charges of these distributions compensate
each other almost completely and their heavy masses are
``absorbed'' by the binding energy. Actually the energy $\pm E(0)$
def\/ines the boundaries of the renormalized electron and positron
zones resulting from the strong polarization of the
electron-positron f\/ield when the excitation appears. But this
excitation could be interpreted as the ``physical'' electron
(positron) if the sequence of the levels in every zone determined
by the vector $\vec P \not= 0 $    were described by the Lorentz
invariant relativistic energy spectrum $\pm E(\vec P)$ of real
particles, that is
\begin{gather}
\label{Eqn47} E(\vec P) = \sqrt{ P^2 + E^2 (0) } = \sqrt{ P^2 +
m^2 }.
\end{gather}

It is worth saying, that the problem of studying of the dynamics
of the self-localized excitation should be solved for any system
with a strong interaction between quantum f\/ields in order to
calculate its ef\/fective mass. For example, a similar problem for
Pekar ``polaron''~\cite{Pekar} in the ionic crystal was considered
in
 \cite{Bogol,Gross,Feynmanpol,PV}  and in a lot of  more recent works. It is
essential that because of the non-linear coupling between the
particle and a self-consistent f\/ield the energy dispersion
$E(\vec P)$ for the quasi-particle proves to be very complicated.
As the result, its dynamics in the crystal is similar to the
motion of the point  ``physical'' particle only at a small enough
total momentum.

However, in the case of QED the problem is formulated in a
fundamentally dif\/ferent way. There is currently no doubt that
the dynamics of the ``physical'' excitation should be described by
the formula~(\ref{Eqn47}) for any(!) values of the momentum $\vec
P$ because of the Lorentz invariance of the Hamiltonian. It means
that the considered nonperturbative approach for describing the
internal structure of the ``physical'' electron should lead to the
energy dispersion law (\ref{Eqn47}) for the entire range of the
momentum $\vec P$.

The rigorous method of taking into account the translational
symmetry in the strong coupling theory for the ``polaron'' problem
was elaborated in the works of Bogoliubov~\cite{Bogol} and
Gross~\cite{Gross}. Let us remind that this method was based on
the introduction of the collective variable  $\vec R$ conjugated
to the total momentum operator  $ \hat {\vec P}$. The canonical
character of the transformation caused by three new variables
$R_i$ was provided by the same number of additional conditions
imposed on the other variables of the system. In the ``polaron''
problem the quantum f\/ield interacting with the particle
contributes to the total momentum of the system. It allows one to
impose these conditions on the canonical f\/ield variables
\cite{Bogol,Gross} and the concrete form of the variable
transformation is based mainly on the permutation relations for
the boson f\/ield operators.

The considered problem has some specif\/ic features in comparison
with the ``polaron'' problem. Firstly, the formation of the
one-particle wave packet is the multi-particle ef\/fect because
this packet includes all initial states of electron-positron
f\/ield as the fermion f\/ield. Secondly, its self-localization is
provided by the polarization potential of the scalar f\/ield that
does not contribute to the total momentum of the system.
Therefore, we use a dif\/ferent approach in order to select the
collective coordinate~$\vec R$. Let us return to the
conf\/iguration representation in the Hamiltonian~(\ref{Eqn18}),
where QED is considered to be the totality of $N$ $( N \rightarrow
\infty)$ point electrons interacting with the quantum
electromagnetic f\/ield in the Coulomb gauge \cite{Heitler}:
\begin{gather}
 \hat H = \sum_{a=1}^{N} \{  \vec{\alpha}_a [ \vecpp 
 + e_0 \hat{ \vec{ A} } (\vec{r}_a)] +
\beta_a m_0 + e_0 \hat \varphi (\vec{r}_a)\} -    \frac{1}{2} \int
d \vec{r}( \vec{\nabla}\hat \varphi (\vec{r}))^2 + \sum_{\vec k
\lambda} \omega (\vec k) \hat n_{\vec k \lambda},
\nonumber\\
\omega (\vec k) = k, \qquad \hat n_{\vec k \lambda} = c^+_{\vec k
\lambda}c_{\vec k \lambda}, \qquad \lambda = 1,2,
\nonumber\\
\hat{ \vec{ A} } (\vec{r}) = \sum_{\vec k \lambda}
\frac{1}{\sqrt{2 k \Omega}} \vec {e}^{(\lambda)} [ c_{\vec k
\lambda} e^{i \vec k \vec r} + c^+_{\vec k \lambda} e^{-i \vec k
\vec r}].\label{10}
\end{gather}

Here $\Omega$ is the normalized volume; $c^+_{\vec k
\lambda}(c_{\vec k \lambda})$ are the operators of the creation
(annihilation) of quanta of a transversal electromagnetic f\/ield,
the quantum having the wave vector $\vec k$, polarization~$\vec
{e}^{(\lambda)}$ and energy $\omega (\vec k) = k$. The sign of the
interaction operators dif\/fers from the standard one because the
parameter $e_0$ is introduced as a positive quantity.

As it was stated above the zero approximation of nonperturbative
QED is def\/ined only by a~strong interaction of electrons with
the scalar f\/ield, where the interaction with the transversal
f\/ield is to be taken into account later in the framework of the
standard perturbation theory. So, the conservation of the total
momentum in the processes with transversal electromagnetic f\/ield
will be provided automatically~\cite{Akhiezer}. Therefore, while
describing the quasi-particle excitation we should consider the
conservation of the total momentum only for the system of
electrons. In the considered representation it is def\/ined by the
sum of the momentum operators of individual particles
\begin{gather*}
 \hat {\vec P} = \sum_{a=1}^{N} \vecpp 
 , \qquad
\hat {\vec P} |  \Phi_1 (\vec P)\rangle  = \vec P |  \Phi_1 (\vec P)\rangle.
\end{gather*}

It means that in the conf\/iguration space the variable $\vec R$
conjugated to the total momentum is simply a coordinate of the
center of mass of the electron-positron system and the desired
transformation to new variables is as follows:
\begin{gather*}
\vec r_a = \vec R + \vec{\rho}_a, \qquad  \vec R =
\frac{1}{N}\sum_{a=1}^{N}\vec r_a, \qquad
\sum_{a=1}^{N} \vec{\rho}_a = 0, \qquad \vec p_a  = - i \vec {\nabla}_a = - \frac{i}{N}\vec {\nabla}_R +  \vec p\,{}'_a , \nonumber\\
  \hat {\vec P} =
-i\vec {\nabla}_R, \qquad  \vec p\,{}'_a  = - i \vec
{\nabla}_{\rho_a} + \frac{i}{N}\sum_{b=1}^{N}\vec
{\nabla}_{\rho_b}, \qquad \sum_{a=1}^{N}  \vec p\,{}'_a  = 0.
\end{gather*}

The Hamiltonian (\ref{10}) with new variables has of the following
form
\begin{gather}
 \hat H = \sum_{a=1}^{N} \left\{  \vec{\alpha}_a \left[- \frac{i}{N}\vec {\nabla}_R + \vecp 
  + e_0 \hat{ \vec{ A} } (\vec{\rho}_a + \vec R)\right] + \beta_a m_0 + e_0 \hat \varphi (\vec{\rho}_a  )\right\}
 \nonumber\\
\phantom{\hat H =}{} - \frac{1}{2} \int d \vec{r}(
\vec{\nabla}\hat \varphi (\vec{r}))^2 + \sum_{\vec k \lambda}
\omega (\vec k) \hat n_{\vec k \lambda}.\label{13}
\end{gather}

It should be noted that the matrix elements of an arbitrary
operator in a new conf\/iguration representation are to be
calculated by integration over the coordinates both of the center
of mass and the relative variables. Let us introduce a special
notation for this norm:
\begin{gather}
\label{14} \langle\langle \Phi_1 | \hat M | \Phi_2 \rangle\rangle
= \int d \vec R \prod_{a} d \vec {\rho}_a \Phi_1^* (\vec R, \{\vec
{\rho}_a\})\hat M \Phi_2 (\vec R, \{\vec {\rho}_a\}).
\end{gather}

The interaction between the ``physical'' electron and the
transversal electromagnetic f\/ield will be taken into account by
means of the perturbation theory (see below Section~\ref{sec5})
and only the scalar f\/ield is considered in the zeroth
approximation. Let us denote by $\hat H_0$ that part of the
operator (\ref{13}) which does not depend on the transversal
electromagnetic f\/ield and describes the internal structure of
the ``physical'' particles in the zero approximation. In fact, the
operator $\hat H_0$ does not depend on the coordinate $\vec R$
because of its commutativity with the operator of the total
momentum of the system of electrons. This also follows from the
well known result~\cite{Heitler} that in the Coulomb gauge the
scalar potential could be excluded from the Hamiltonian. As a
result the operator $\hat H_0$ depends only on the vector
dif\/ferences $(\vec {r}_a - \vec {r}_b) = (\vec {\rho}_a - \vec
{\rho}_b)$ and does not change with the simultaneous translation
of all the coordinates. As a consequence, the eigenfunctions of
the Hamiltonian $\hat H_0$ depend on the coordinate $\vec R$ in
the same way as for a free particle:{\samepage
\begin{gather*}
  \Phi (\vec R, \{\vec {\rho}_a\}) = \frac{1}{(2\pi)^{3/2}}e^{i\vec P\vec R} |\Phi_1 (\vec
P, \{\vec {\rho}_a\})\rangle,\nonumber\\
\hat H_0 \rightarrow \hat H_0 (\vec P) = \sum_{a=1}^{N} \left\{
\vec{\alpha}_a \left[ \frac{1}{N}\vec {P} +
\vecp \right] + \beta_a m_0 + e_0 \hat \varphi (\vec{\rho}_a
)\right\} -  \frac{1}{2} \int d \vec{r}( \vec{\nabla}\hat
\varphi (\vec{r}))^2.
\end{gather*}}

Further calculations consist in returning to the f\/ield
representation by the variables   $ \vec {\rho}_a$ in the limit $(
N \rightarrow \infty)$ and in using the approximate trial wave
packet $|\Phi_1 (\vec P)\rangle$ similar to (\ref{Eqn20}) but with
the coef\/f\/icient functions depending on  $\vec P$. Thus, in the
framework of non-perturbation QED the orthogonal  and normalized
set of states for the   one-particle excitation of the
electron-positron f\/ield is def\/ined as follows:
\begin{gather}
\label{16} |\Phi^{(0)}_1(\vec P) \rangle \simeq
\frac{1}{(2\pi)^{3/2}}e^{i\vec P\vec R} \!\int d \vec{q} \{
U_{\vec{q}s}(\vec P) a^+_{\vec{q} s} + V_{\vec{q}s}(\vec P)
b^+_{\vec{q} s} \} e^{i \vec q \cdot \vec \rho}| 0; 0;\varphi(\vec
\rho)\rangle,  \qquad \vec r = \vec R + \vec \rho,\!\!
\end{gather}
with the norm (\ref{14}) and the coef\/f\/icient functions
$U_{\vec{q}s}(\vec P)$, $V_{\vec{q}s}(\vec P)$.

Using the coordinate representation for these functions
\begin{gather}
 \Psi_{\nu} (\vec r, \vec P) = \int \frac{d \vec q}{(2\pi)^{3/2}} \sum_{s} U_{\vec{q}s}(\vec P)
u_{\vec{q} s \nu} e^{i \vec q \vec r}, \qquad \Psi^c_{\nu} (\vec
r) = \int \frac{d \vec q}{(2\pi)^{3/2}} \sum_{s} V_{\vec{q}s}(\vec
P) v_{\vec{q} s \nu} e^{i \vec q \vec r},\label{Eqn48}
\end{gather}
one can f\/ind the following functional for calculating the value
$E(\vec P)  $ corresponding to the energy of one-particle
excitation with an arbitrary momentum
\begin{gather}
 E(\vec P) = \int d \vec r \Big\{ \Psi^+ (\vec r, \vec P) \Big[(\vec \alpha \vec P -i\vec \alpha \vec \nabla
+ \beta m_0) + e_0\frac{1}{2} \varphi (\vec r, \vec P) \Big] \Psi
(\vec r, \vec P)
\nonumber\\
\phantom{E(\vec P) =}{}- \Psi^{+c} (\vec r, \vec P) \big[(- \vec
\alpha \vec P -i\vec \alpha \vec \nabla + \beta m_0) +
e_0\frac{1}{2} \varphi (\vec r, \vec P) \Big]\Big\} \Psi^c (\vec r, \vec P), \nonumber\\
\varphi (\vec r, \vec P) = e_0 \int \frac {d \vec r\,{}'} {|\vec r
- \vec r\,{}'|} [\Psi^{+} (\vec r\,{}', \vec P) \Psi (\vec r\,{}',
\vec P) - \Psi^{+c} (\vec r\,{}', \vec P) \Psi^{c} (\vec r\,{}',
\vec P)],
\nonumber\\
\int {d \vec{r}} \, [\Psi^{+} (\vec r, \vec P) \Psi (\vec r, \vec
P) + \Psi^{+c} (\vec r\,{}', \vec P) \Psi^{c} (\vec r\,{}', \vec
P)] = 1.\label{Eqn49}
\end{gather}

It should be noted that the mean value of the operator $\hat H_0
(\vec P)$ in the representation of the second quantization was
calculated by taking into account the expression
\[
\lim_{N \to \infty} \frac{1}{N} \sum_{\vec p} = 1,
\]
that follows from a well known formula for the density of states
when EPF is quantized within the normalized volume $\Omega$
\cite{Akhiezer}.

It was mentioned above, that there is quite a close analogy
between the nonperturbative description of QED and the theory of
strong coupling in the ``polaron'' problem. Therefore, it is of
interest to compare the obtained functional (\ref{Eqn49}) with the
results of various methods of including the translational motion
in the ``polaron'' problem. The simplest one was used by Landau
and Pekar~\cite{PV} who introduced the Lagrange multipliers in the
form
\begin{gather}
\label{Eqn49a} J(\vec P) = J( \vec P =0) + (\vec P \vec V),
\end{gather}
with the functional $J( \vec P =0)$ referring to a resting
``polaron'' and the Lagrange multiplier  $V_i$ denoting the
components of the quasi-particle average velocity.

We can see that the obtained functional~(\ref{Eqn49}) has the same
form as  (\ref{Eqn49a}) if the relativistic velocity of the
excitation is determined by the formula
\begin{gather*}
\vec V = \int d \vec r \{\Psi^+ (\vec r, \vec P) (\vec \alpha)
\Psi (\vec r, \vec P) + \Psi^{+c} (\vec r, \vec P) (\vec \alpha)
\Psi^c (\vec r, \vec P)\} ,
\end{gather*}
corresponding to the well known interpretation of  Dirac matrices
\cite{Akhiezer}.

Varying this functional by taking into account the normalized
conditions leads to the following equations for the wave functions
\begin{gather}
 \{(\vec \alpha \vec P -i\vec \alpha \vec \nabla + \beta m_0) + e_0 \varphi (\vec r, \vec P) -
E(\vec P)\} \Psi (\vec r, \vec P) = 0,
\nonumber\\
\{(\vec \alpha \vec P + i\vec \alpha \vec \nabla - \beta m_0) -
e_0 \varphi (\vec r, \vec P) - E(\vec P) \} \Psi^c (\vec r, \vec
P)= 0.\label{Eqn50}
\end{gather}

These equations show that  unlike the ``polaron'' problem
\cite{Bogol}, the translational motion of the quasi-particle
determining the momentum $\vec P$ is related in our case to its
internal movement described by the coordinate $\vec r$  by means
of spinor variables only. The physical reason for this separation
of variables is explained by the fact that in QED the
self-localized state is formed by the scalar f\/ield, and its
interaction with the particle does not involve the momentum
exchange. In order to f\/ind the analytical energy spectrum
$E(\vec P)$ the system of non-linear equations~(\ref{Eqn50})
should be diagonalized with respect to to the spinor variables.
The possibility of such diagonalization seems to be a non-trivial
requirement for the nonperturbative QED under consideration.

The solution of the equations  (\ref{Eqn50}) can be found on the
basis of the states for which the dependence on the vector $\vec
q$ in the wave packet amplitudes  $U_{\vec q, s}$, $V_{\vec q, s}$
remains the same as it was in the motionless ``physical''
electron. However, the relation between the spinor components of
these functions can be changed. But as the self-consistent scalar
potential involves the summation over all spinor components it
does not depend on the momentum for the class of states in
question:
\begin{gather}
\label{Eqn51} \varphi (\vec r, \vec P) = \varphi (r) |_{\vec P =
0}.
\end{gather}

In the coordinate representation the transformation of the spinor
components of the wave functions satisfying the equations
(\ref{Eqn50}) takes place because of the dependence on the
momentum. The solution can be constructed by sorting out various
linear combinations of the wave functions $\Psi (\vec r)$, $\Psi^c
(\vec r)$, $\tilde{\Psi}(\vec r)$, $\tilde{\Psi^c} (\vec r) $
found in Section~\ref{sec2}  for a resting electron. These
functions correspond to the degenerated states in the case of
$\vec P = 0$ but are mixed for a moving electron. It is found that
there is only one normalized linear combination satisfying all the
necessary conditions of self-consistency:
\begin{gather}
 \Psi (\vec r, \vec P)  = L (\vec P) \Psi (\vec r) + K (\vec P) \tilde{\Psi^c} (\vec r) ,
\qquad \Psi^c(\vec r, \vec P)  = L_1(\vec P) \Psi^c (\vec r) +
K_1(\vec P) \tilde{\Psi}(\vec r), \nonumber\\
|L|^2 + |K|^2 = |L_1|^2 + |K_1|^2 = 1.\label{Eqn52}
\end{gather}

The condition  (\ref{Eqn51}), according to which the potential
does not depend on the momentum for the excitation at the energy
$E(\vec P)$, is fulf\/illed if the coef\/f\/icients are related as
\begin{gather*}
\label{Eqn52a} L_1 = - K, \qquad K_1 = L.
\end{gather*}

These relations are also consistent with  equations (\ref{Eqn50})
for  wave functions.

This means that the ``physical'' electron moves in such a way that
its states are transformed in the phase space of the orthogonal
wave functions (\ref{Eqn28a})--(\ref{Eqn33b}) but the amplitudes
of its ``internal'' charge distributions are not changed. The
analogous approach is considered for Lorentz invariant
transformation of the bispinors corresponding to the free
particles in QED \cite{Drell}.

Substituting the superpositions  (\ref{Eqn52}) into equations
(\ref{Eqn50}) we use the following relations:{\samepage
\begin{gather*}
 (\vec \alpha \vec P) \Psi = \left( \begin{array}{c}
i f (r) (\vec \sigma \vec P)\Omega_{1/2, 1, M}\\
g(r) (\vec \sigma \vec P) \Omega_{1/2,0,M}
\end{array} \right), \qquad
(\vec \alpha \vec P) \Psi^c = \left( \begin{array}{c}
g_1 (r) (\vec \sigma \vec P)\Omega_{1/2, 1, M}\\
-i f_1(r) (\vec \sigma \vec P) \Omega_{1/2,0,M}
\end{array} \right),\nonumber\\
(- i \vec \alpha \vec \nabla + \beta m_0 + e_0 \varphi) \Psi =
E(0) \left( \begin{array}{c}
g (r) \Omega_{1/2, 0, M}\\
i f(r) \Omega_{1/2,1,M}
\end{array} \right), \nonumber\\
(-i \vec \alpha \vec \nabla + \beta m_0 + e_0 \varphi)  \Psi^c = -
E(0) \left( \begin{array}{c}
-i f_1 (r) \Omega_{1/2, 0, M}\\
g_1(r) \Omega_{1/2,1,M}
\end{array} \right),
\end{gather*}
and similar formulas for the functions  $\tilde{\Psi}(\vec r)$,
$\tilde{\Psi^c}(\vec r)$,  $\sigma_i$ are the Pauli matrices.}

For equations  (\ref{Eqn50}) to be fulf\/illed for any vector
$\vec r$ it is necessary to set the coef\/f\/icients of spherical
spinors equal to the same indexes  $l$. The corresponding radial
functions are proved to be the same under these conditions, and
the following system of equations for the spinors
$\chi_{0,1}^{\pm}$ is obtained
\begin{gather}
 i L (\vec \sigma \vec P)\chi^+_1 + K ( E + E_0)\chi_1^+ = 0, \qquad i K (\vec \sigma \vec P)\chi^+_0
+ L(
E - E_0)\chi_0^+ = 0,\nonumber\\
L (\vec \sigma \vec P)\chi^+_0 + i K ( E + E_0)\chi_0^+ = 0,
\qquad  K (\vec \sigma \vec P)\chi^+_1 + i L(
E - E_0)\chi_1^+ = 0,\nonumber\\
L_1 (\vec \sigma \vec P)\chi^-_1 - i K_1 ( E - E_0)\chi_1^- = 0,
\qquad  K_1 (\vec \sigma \vec P)\chi^-_0 - i L_1(
E + E_0)\chi_0^- = 0,\nonumber\\
i L_1 (\vec \sigma \vec P)\chi^-_1 -  K_1 ( E - E_0)\chi_1^- = 0,
\qquad i K_1 (\vec \sigma \vec P)\chi^-_0 -L_1( E + E_0)\chi_0^- =
0,\label{Eqn54}
\end{gather}
\noindent where the notation $E(0) \equiv E_0$  was used.

Spin variables in equations (\ref{Eqn54}) are also separated. In
order to show this one can use, for example, the relation between
the coef\/f\/icients resulting from the 4th equation in
(\ref{Eqn54}) in the f\/irst one of these equations:
\[
\chi_1^+ = i \frac{K (\vec \sigma \vec P)\chi^+_1}{L( E - E_0)}.
\]

As a result there exists a non-trivial solution of these equations
for two branches of the energy spectrum
\begin{gather}
\label{Eqn55} E_{e,p} = \pm \sqrt{E^2_0 + P^2},
\end{gather}
referring to the electron and positron zones, respectively
\cite{SLS}. The same expressions can be obtained for all
conjugated pairs of the equations in~(\ref{Eqn54}). The
coef\/f\/icients in the wave functions~(\ref{Eqn52}) can be found
taking into consideration the normalization condition:
\begin{gather}
 L^e = K^e_1 = \frac{P}{\sqrt{P^2 + (E_e - E_0)^2}}, \qquad K^e = - L^e_1 =
\frac{E_e - E_0}{\sqrt{P^2 + (E_e - E_0)^2}}, \nonumber\\
L^p = K^p_1 = \frac{P}{\sqrt{P^2 + (E_e + E_0)^2}}, \qquad K^p = -
L^p_1 = -\frac{E_e + E_0}{\sqrt{P^2 + (E_e  +
E_0)^2}}.\label{Eqn56}
\end{gather}

One can see that this set of coef\/f\/icients coincides with the
set of spinor components for solving the Dirac equation for a free
electron with the observed mass $ m = E_0$ \cite{Drell}. Thus, the
results of this section show that the ``internal'' structure of
the ``physical'' electron (positron) considered in this paper is
consistent with the energy dispersion (\ref{Eqn55}) for a real
free particle due to the relativistic invariance of the Dirac
equation.

For the interpretation of the wave packet (\ref{16}) as the state
vector for a ``physical'' particle it is essential that it is the
eigenvector of the total momentum of the electron-positron f\/ield
in the framework of the considered zeroth approximation. It means
that the ``physical'' electrons with dif\/ferent momenta form the
set of orthogonal and normalized functions if the condition
(\ref{14}) is taken into account:
\begin{gather*}
\langle\langle \Phi^{(0)}_1 (\vec P_1) |  \Phi^{(0)}_1 (\vec
P)\rangle\rangle = \delta (\vec P - \vec P_1).
\end{gather*}

\section[Perturbation theory for QED\\ with the ``physical'' electron-positron field. Discussion]{Perturbation theory for QED\\ with the ``physical'' electron-positron f\/ield. Discussion}\label{sec5}

It is well known that the renormalizability is one of the most
important features of QED and it is conf\/irmed by the coincidence
of its results with the experimental data.  Actually it means that
the calculated characteristics of the real electromagnetic
processes do not depend on the initial parameters  $e_0$, $m_0$
but only on the observed values of $e$ and $m$. So, we should
clearly show that the ef\/fects of the nonperturbative QED refer
to the internal structure of the ``physical'' electron (positron)
only. However, its interaction with real electromagnetic f\/ield
should be def\/ined by the standard perturbation theory with the
f\/ine structure constant $\alpha \ll 1$.

Besides, it should be demonstrated that the integrals which
def\/ine the corrections to the zero approximation in the
perturbation theory with charge $e \sim e^{-1}_0$ are converged
without introducing any additional regularizating parameters
including the cut-of\/f momentum.

To solve these problems,  it is necessary to consider the form of
the perturbation theory which uses the basis of states
corresponding to the ``physical'' electron (positron) with various
momentum.     Taking into account the results of the last section
let us represent the QED Hamiltonian (\ref{Eqn18}) in the
following form:
\begin{gather*}
\hat H = \hat H_0 + \hat H_I, \nonumber\\
\hat H_0 = \int d \vec{r}\left\{ : \!\hat \psi^* (\vec{r}) [ -
i\vec \alpha ( \vec {\nabla}_{\vec R} + \vec {\nabla}_{\vec r}) +
\beta m_0] \hat \psi (\vec{r})\!: + e_0 \hat \varphi (\vec{r})
:\!\hat \rho (\vec{r})\!:  - \frac{1}{2} ( \vec{\nabla}\hat
\varphi
(\vec{r}))^2\right\} \nonumber\\
\phantom{\hat H_0 =}{} + \sum_{\vec k \lambda} \omega (\vec k) \hat n_{\vec k \lambda},\nonumber\\
\hat H_I = e_0 \int d \vec{r} :\! \hat \psi^* (\vec{r}) [\vec
\alpha \hat{ \vec{ A} } (\vec{r} + \vec{R})] \hat
\psi (\vec{r})\!:.
\end{gather*}

The fact that the Hamiltonian depends on the coordinate $\vec R$
canonically conjugated to the total  momentum of the
electron-positron f\/ield  determines the spontaneous breaking of
symmetry of the system. This situation was discussed by Bogoliubov
\cite{quasi} for a lot of concrete physical systems. However, the
global symmetry of the system   is reconstructed when the observed
values are averaged by the coordinate  $\vec R$ in accordance with
the norm~(\ref{14}).

In the previous sections the spectrum of the one-particle
excitations of the electron-positron f\/ield was calculated
approximately in the leading order in a power series of
$\alpha_0^{-1}$. Let us now consider the matrix elements of the
perturbation operator $H_I$ that are def\/ined by the transitions
between the states corresponding to the ``physical'' electrons
(positrons) with the momenta   $\vec P$ and $\vec P_1$ and with
emission (absorption) a quanta of the transversal electromagnetic
f\/ield. It is important that the vacuum state of the ``physical''
electron-f\/ield is also renormalized with respect to to the
``primary'' vacuum of the ``bare'' particles. Actually this new
vacuum is the wave function of the collective quasi-particle
states localized in the vicinity of various coordinates~$\vec R$.

Let us write the obvious form of the initial $ | i \rangle $ and
f\/inal $ | f \rangle $  states of this process using the
def\/inition (\ref{16}) for the one-particle wave packet and the
state vector of the free electromagnetic f\/ield corresponding to
$N$ quanta with the wave vector $\vec k$ and polarization
$\lambda$:
\begin{gather*}
 | i \rangle =  |\Phi_1(\vec P), \Phi_1 (0){N^{(i)}_{\vec k,\lambda}} \rangle  = \frac{1}{(2\pi)^{3/2}}e^{i\vec
P\vec R } \int d \vec{q} \{ U_{\vec{q},s}(\vec P) a^+_{\vec{q} s}
+ V_{\vec{q},s}(\vec P) b^+_{\vec{q} s} \}e^{i\vec q
\cdot(\vec r - \vec R )}  \nonumber\\
\phantom{| i \rangle =}{} \times\frac{1}{(2\pi)^{3/2}} \! \int\! d
\vec{q} \{ U_{\vec{q},s}(0) a^+_{\vec{q} s} + V_{\vec{q},s}(0)
b^+_{\vec{q} s} \}e^{i\vec q \cdot(\vec r - \vec R_1)}
\frac{(c^+_{\vec k, \lambda})^{N^{(i)}_{\vec
k,\lambda}}}{\sqrt{N^{(i)}_{\vec
k,\lambda}!}}  |0; 0;\varphi(\vec r - \vec R ); \varphi(\vec r - \vec R_1)\rangle, \nonumber\\
| f \rangle =  |\Phi_1(0), \Phi_1 (\vec P_1){N^{(f)}_{\vec
k,\lambda}} \rangle  = \frac{1}{(2\pi)^{3/2}}e^{i\vec P_1\vec R_1}
\int d \vec{q} \{ U_{\vec{q},s}(\vec P_1) a^+_{\vec{q} s} +
V_{\vec{q},s}(\vec P_1) b^+_{\vec{q} s} \}e^{i\vec q
\cdot(\vec r - \vec R_1 )} \\ 
\phantom{| f \rangle =}{} \times\frac{1}{(2\pi)^{3/2}} \! \int\! d
\vec{q} \{ U_{\vec{q},s}(0) a^+_{\vec{q} s} + V_{\vec{q},s}(0)
b^+_{\vec{q} s} \}e^{i\vec q \cdot(\vec r - \vec R )}
\frac{(c^+_{\vec k, \lambda})^{N^{(f)}_{\vec
k,\lambda}}}{\sqrt{N^{(f)}_{\vec k,\lambda}!}} |0; 0;\varphi(\vec
r - \vec R ); \varphi(\vec r - \vec R_1)\rangle. \nonumber
\end{gather*}

Let us calculate the matrix element of the operator $\hat H_I$ for
the transition of the ``physical'' electron between the states
with 4-momenta   $P = (\vec P, E)$ and $P_1 = (\vec P_1, E_1)$
with the emission of one quantum of the transversal
electromagnetic f\/ield with   $ k =(\vec k, \omega)$. It is
supposed that both fermion and photon 4-momenta are out of the
mass surface. In this case the matrix element    $M_{fi} = \Gamma
(P, k)\delta( P - P_1 - k)$ def\/ines the vertex function $\Gamma
(P, k)$ in the considered representation. The norm
 (\ref{14}) is used for calculation:
\begin{gather*}
 M_{fi} =   \langle \langle f |\hat H_I |i \rangle \rangle  = \frac{e_0}{\sqrt{2 k \Omega}}\sum_{s,s'} \sum_{\mu,\nu}\int d
\vec{r}\int d \vec{R}\int d \vec{R_1} e^{i (\vec P\cdot \vec R -
\vec P_1\cdot \vec R_1)} e^{- i  \vec k\cdot
\vec r }\nonumber\\
\phantom{M_{fi} =}{} \times \int
 \frac{ d \vec{q}}{(2\pi)^{3/2}}
\int \frac{ d \vec{q}\,{}'}{(2\pi)^{3/2}}  \big\{
\big[U^*_{\vec{q}\,{}',s'}(\vec{P}_1)U
_{\vec{q}\,{}',s'}(0)U^*_{\vec{q},s}(0)
U_{\vec{q},s}(\vec{P}) \nonumber\\
\phantom{M_{fi} =}{} - V^*_{\vec{q}\,{}',s'}(\vec{P}_1)V
_{\vec{q}\,{}',s'}(0)V^*_{\vec{q},s}(0)
V_{\vec{q},s}(\vec{P})\big]  \nonumber\\
\phantom{M_{fi} =}{}\times\big[u^*_{\vec{q}s\mu} (\vec {\alpha}
\vec {e}^{\lambda})_{\mu,\nu}u_{\vec{q}s\nu} +
u^*_{\vec{q}\,{}'s'\mu} (\vec {\alpha} \vec
{e}^{\lambda})_{\mu,\nu}u_{\vec{q}\,{}'s'\nu}\big]\big\} e^{i(\vec
r - \vec R)\cdot \vec q} e^{i(\vec r - \vec R_1)\cdot \vec
{q}\,{}'}.
\end{gather*}

If the coordinate representation  (\ref{Eqn48}) is used for the
coef\/f\/icient functions $U_{\vec{q},s}$, $V_{\vec{q},s}$, the
result is:
\begin{gather}
 M_{fi} =   \frac{e_0}{\sqrt{2 k \Omega}}\int \! d \vec{r} \! \int \! d \vec{R}\! \int \! d \vec{R_1}\big\{\big[ \Psi^*
(\vec r - \vec R , 0) (\vec {\alpha} \vec {e}^{\lambda})\Psi
(\vec r -\vec {R}, \vec P)\Psi^* (\vec r - \vec
R_1,\vec P_1) \Psi  (\vec r -\vec {R}_1, 0)  \nonumber\\
\phantom{M_{fi} = }{} +\Psi^* (\vec r - \vec R , 0)  \Psi (\vec r
-\vec {R}, \vec P)\Psi^* (\vec r - \vec R_1,\vec P_1)(\vec
{\alpha} \vec {e}^{\lambda}) \Psi  (\vec r -\vec {R}_1, 0)\big]
\nonumber\\
\phantom{M_{fi} = }{}- \big[ \Psi^{c,*} (\vec r - \vec R , 0)
(\vec {\alpha} \vec {e}^{\lambda})\Psi^c  (\vec r -\vec {R}, \vec
P)\Psi^{c,*} (\vec r - \vec R_1,\vec P_1) \Psi^c  (\vec r -\vec {R}_1, 0) \label{Eqn59a} \\
\phantom{M_{fi} = }{} +\Psi^{c,*} (\vec r - \vec R , 0)  \Psi^c
(\vec r {-}\vec {R}, \vec P)\Psi^{c,*} (\vec r {-} \vec R_1,\vec
P_1)(\vec {\alpha} \vec {e}^{\lambda}) \Psi^c (\vec r {-}\vec
{R}_1, 0)\big]   \big\} e^{i (\vec P\cdot \vec R {-} \vec P_1\cdot
\vec R_1)} e^{- i \vec k \vec r} \!.\nonumber
\end{gather}

Let us pass to the integration over the coordinate of the center
of mass $\vec R_c = (\vec R + \vec R_1)/2$ and relative coordinate
$\vec R_0 = (\vec R - \vec R_1)$ in equation~(\ref{Eqn59a}) and
introduce the following vectors:
\[
\vec r_1 = \vec r - \vec R_c - \frac{1}{2}\vec R_0, \quad \vec r_2
= \vec r - \vec R_c + \frac{1}{2}\vec R_0.
\]

Then the vertex function results in:
\begin{gather}
 M_{fi} = \Gamma ( P,   k) \delta (\vec P - \vec P_1 - \vec k), \qquad
  P_1 =  P -   k,\nonumber\\
\Gamma (\vec P, \vec k) = \frac{e_0}{\sqrt{2 k \Omega}}\int d \vec
r_2 e^{i  \vec P_1 \cdot \vec r_2 } \int d \vec r_1  e^{- i \vec P
\vec r_1}\big\{ \Psi^* (r_2, \vec {P}_1) (\vec {\alpha} \vec
{e}^{\lambda})\Psi
(\vec r_2, 0)\Psi^* (\vec r_1 , 0)  \Psi (\vec r_1, \vec {P})  \nonumber\\
\phantom{\Gamma (\vec P, \vec k) =}{} +\Psi^* (\vec r_2, \vec
{P}_1) \Psi (\vec r_2, 0)\Psi^* (\vec r_1, 0)(\vec {\alpha} \vec
{e}^{\lambda})
\Psi (\vec r_1, \vec {P}) \nonumber\\
\phantom{\Gamma (\vec P, \vec k) =}{} - \Psi^{c,*}  (\vec r_2,
\vec {P}_1) (\vec {\alpha} \vec {e}^{\lambda})\Psi^c  (\vec  r_2,
0)\Psi^{c,*} (\vec  r_1 , 0)  \Psi^c (\vec  r_1, \vec {P})
\nonumber\\
\phantom{\Gamma (\vec P, \vec k) =}{}-\Psi^{c,*}  (\vec  r_2, \vec
{P}_1)  \Psi^c  (\vec  r_2, 0)\Psi^{c,*} (\vec  r_1 , 0)(\vec
{\alpha} \vec {e}^{\lambda})  \Psi^c (\vec  r_1, \vec {P})
\big\}.\label{Eqn59c}
\end{gather}

The linear combinations (\ref{Eqn52}) and the wave functions
$\Psi$, $\Psi^c$, $\tilde{\Psi}$, $\tilde{\Psi}^c$ from equations
(\ref{Eqn28a}), (\ref{Eqn31b}), (\ref{Eqn33b}), (\ref{Eqn33c}) for
the   resting ``physical'' electrons should be used in
equation~(\ref{Eqn59c}). Then one can pass to the two-component
spinors in the integrand and f\/ind the following formula:
\begin{gather*}
 \Gamma   =  i\frac{e_0}{\sqrt{2 k \Omega}}\int d \vec r_1  \int d \vec r_2 \big\{  \rho(\vec r_1)\rho
(\vec r_2) (\chi^{+*}_0 (\vec {\sigma}
 \vec {e}^{\lambda})\chi^+_0)[K^* (\vec {P}_1,E_1) L (\vec {P},E) \nonumber\\
 \phantom{\Gamma   = }{} - L^* (\vec {P}_1,E_1) K (\vec {P },E)]
- \rho_1 (\vec r_1)\rho_1 (\vec r_2) (\chi^{-*}_0 (\vec {\sigma}
 \vec {e}^{\lambda})\chi^-_0)  \nonumber\\
\phantom{\Gamma   = }{} \times [K_1^* (\vec {P}_1,E_1) L_1 (\vec
{P},E) - L_1^* (\vec {P}_1,E_1) K_1 (\vec {P},E)]\big\}
 e^{i \vec P_1 \vec r_1 } e^{-i \vec P \vec r  },
\end{gather*}
 where the functions $L (\vec {P},E)$, $K (\vec {P },E)$, $L_1 (\vec {P},E)$, $K_1 (\vec {P},E) $ are def\/ined by
equations (\ref{Eqn56}) but with the energy variable out of the
mass surface $E \neq  E_0$. The charge densities are def\/ined as
follows:
\[
\rho(\vec r) = g^2(r) + f^2(r) , \qquad \rho_1(\vec r) = g^2_1(r)
+ f^2_1(r).
\]

Accordingly to Section~\ref{sec2} the functions $f$, $g$ and
$f_1$, $g_1$, that describe the ``physical'' electron internal
structure, can be expressed over the universal function and are
dif\/fered from each other only by the normalization conditions
\begin{gather*}
g(r) = \frac{1}{1+C}m_0^{3/2}\frac{u_0 (r m_0)}{r}, \qquad g_1(r)
= \frac{C}{1+C}m_0^{3/2}\frac{u_0 (r m_0)}{r},
\\
f(r) = \frac{1}{1+C}m_0^{3/2}\frac{v_0 (r m_0)}{\rho}, \qquad
g_1(r) = \frac{C}{1+C}m_0^{3/2}\frac{v_0 (r m_0)}{r}.
\end{gather*}

Besides, the spin and angular variables can be excluded by means
of equations~(\ref{Eqn52}) for the spinor functions of the
``physical'' electron. As for example,
\[
K (\vec {P})(\vec {\sigma} \vec {e}^{\lambda})\chi^+_0  = \frac{iL
(\vec {P})}{E + E_0}(\vec {\sigma} \vec {e}^{\lambda})(\vec
{\sigma} \vec {e}^{\lambda})\chi^+_0 =\frac{iL (\vec {P})}{E +
E_0}(\vec {P} \vec {e}^{\lambda})\chi^+_0.
\]

Taking into account the normalization conditions for the spinors
$\chi^{+,- }_{0}$ one can f\/ind:
\begin{gather}
 \Gamma (  P,  k) =  \frac{e_0}{\sqrt{2 k \Omega}}\frac{(1-C)}{(1+C)} L (\vec {P},E)L (\vec
{P}_1,E_1)\left[\frac{(\vec {P} \vec {e}^{\lambda})}{E  + E(0)}  -
\frac{(\vec {P}_1 \vec {e}^{\lambda})}{E_1 + E(0)}\right]
 F^*(|\vec P - \vec k|)F(|\vec P|)], \nonumber\\
F(|\vec P|) \equiv F(\alpha, |\vec P|/m) =  \int_0^{\infty} dx
[u^2_0(x) +  v^2_0(x)] \frac{\sin |\vec P| x/m_0}{|\vec P| x/m_0}
\nonumber\\
\phantom{F(|\vec P|)}{}   = \int_0^{\infty} dx [u^2_0(x) +
v^2_0(x)] \frac{\sin \alpha |\vec P| x/2 |a_0| m}{\alpha |\vec P|
x/2 |a_0| m}.\label{Eqn63}
\end{gather}

If one use the relations $e_0 (1 - C)/(1 + C) = e$, $(1 - C)/(1 +
C)\sim e_0^{-2}$, $E(0) = m$ and put $F(|\vec P|) = 1$ the
expression (\ref{Eqn63}) will coincide completely with the
analogous matrix element in standard form of QED with the
renormalized charge $e$ and mass $m$ of the
electron~\cite{Heitler}. It means that the series for the observed
characteristics of the ``physical'' electron in a reverse power of
the ``primary'' charge $e_0^{-1}$ leads to the terms with the same
structure as the standard perturbation theory in a power of $e$.

However, equation~(\ref{Eqn63}) shows that behavior of the
renormalized vertex function $\Gamma ( P,k)$ at large arguments is
def\/ined by the Fourier transformation of the radial functions
that describe the charge distribution in the ``physical''
electron. Analysis of the asymptotic behavior of the solutions of
equations~(\ref{36}) shows that the form-factor $F(|\vec P|)$ at
large   $ |\vec P| \gg m_0$ is described as follows:
\begin{gather}
 F(|\vec P|) \sim  \frac{L_0^4}{(|\vec P|^2 +   L_0^2)^2}, \qquad
L_0 = 2 m_0 = m  \frac{4 |a_0|}{\alpha},\label{Eqn64}
\end{gather}
and tends to zero rather quickly.

So, if the representation with the basis of states corresponding
to the ``physical'' electrons was used in the zeroth approximation
then the cut-of\/f momentum $L_0 \sim 1$~GeV  appears naturally
and it has def\/inite physical meaning. It means that in this
representation the perturbation theory in a power of the observed
value of the f\/ine structure constant $\alpha \ll 1$ does not
contain the ultraviolet divergencies in the integrals.

It may seem at f\/irst sight that the f\/inite value of the
cut-of\/f momentum is in the contradiction with the experiment. It
is well known that the comparison of the theoretical calculations
in the framework of the standard renormalized QED with the
experimental data even in the low energy range (for example, for
anomalous magnetic moment) shows that no additional momentum $L
\lesssim  10^7$~GeV  can be introduced in QED (for example,
\cite{Brodsky}). However, these results do not maintain  that the
electron is the point-like object with the form-factor equal to
unity in the entire range of the momentum because   the
form-factor of the electron appears due to the high order
corrections of the perturbation theory with any calculation
scheme~\cite{Akhiezer}. Actually it is followed from these results
that the electron cannot be considered as the composite system
including any parameter $L$, as for example, the hypothetical
particle with a mass~$ M \sim L$~\cite{Brodsky}.

But the most essential feature of the present approach is the
result that the cut-of\/f momentum~$L_0$ is not the additional
parameter of the theory because it can be expressed only in terms
of the observed values of QED, namely $\alpha$ and $m$(!). The
form factor of the ``physical'' electron in this case can be
considered as the result of the partial summation of high-order
corrections of usual perturbation series. Let us now show that the
appearance of $L_0$ at the intermediate calculation stages may not
contradict to the fact that it is not ef\/fect at the f\/inal
results for the observed physical values in the renormalised QED.

Really, let us consider a general structure of the
non-renormalized form factor of the electron as the function of
the invariant parameters in the framework of QED with an arbitrary
cut-of\/f momentum $L$ and the non-renormalized values of
$\alpha_0$, $m_0 $. For def\/initeness, let us write only its
scalar part:
\begin{gather}
\label{65} F_0(k) \equiv F_0 \left(\alpha_0; \frac{k^2}{m_0^2};
\frac{m_0 }{L }\right).
\end{gather}

Computation of this form factor by means of the perturbation
theory with the standard renormalization scheme is def\/ined by
the calculation of the vertex function   \cite{Brodsky} and
corresponds to the expansion of the function   (\ref{65}) as a
power series in both coupling constant $\alpha_0$ and the
parameter $m_0 /L $:
\begin{gather}
 F_0(k) \approx 1 +  \alpha_0 \left[ F^{(0)}_{0 1} \left(0; \frac{k^2}{m_0^2};0\right) + \frac{m_0 }{L } F^{(1)}_{0 1}
\left(0; \frac{k^2}{m_0^2};0\right)  + \cdots\right]  \nonumber\\
\phantom{F_0(k) \approx} +\alpha^2_0 \left[ F^{(0)}_{0 2} \left(0;
\frac{k^2}{m_0^2};0\right) + \frac{m_0 }{L } F^{(1)}_{02} \left(0;
\frac{k^2}{m_0^2};0\right) + \cdots\right] + \cdots.\label{66}
\end{gather}

Let us remind that the QED renormalizability means that all
coef\/f\/icients of these series are remained f\/inite in the
limit $L \rightarrow \infty $ if the parameters $\alpha_0$, $m_0 $
are changed by their renormalized values $\alpha$, $m$. Then the
theoretically calculated value of the form factor is def\/ined by
the expansion in a power series of both $\alpha$ and an
independent parameter   $\xi \sim m/L$:
\begin{gather}
 F_{\rm th}(k) \approx 1 +  \alpha \tilde{F}^{(0)}_{0 1} \left(  \frac{k^2}{m^2} \right) + \alpha^2 \tilde{
F}^{(0)}_{0 2} \left(  \frac{k^2}{m ^2} \right)+ \cdots \nonumber\\
\phantom{F_{\rm th}(k) \approx}{} + \frac{m}{L } \left[\alpha
\tilde{F}^{(1)}_{0 1} \left( \frac{k^2}{m ^2} \right) + \alpha^2
\tilde{F}^{(1)}_{02} \left(  \frac{k^2}{m ^2} \right) +
\cdots\right] + \cdots.\label{67}
\end{gather}

The formal series that  appear in the intermediate stages and
def\/ine the relations between the ``primary'' parameters of the
theory   $\alpha_0$, $m_0 $ and the observed value   $\alpha$, $m$
can be represented as follows:
\begin{gather}
 \alpha \approx \alpha_0 A_1\left(\frac{m_0 }{L }\right) +  \alpha_0^2 A_2\left(\frac{m_0 }{L }\right) + \cdots = \alpha_0
Z\left(\alpha_0,\frac{m_0 }{L }\right), \nonumber\\
m \approx m_0 \left[1 + \alpha_0 B_1\left(\frac{m_0 }{L }\right) +
\alpha_0^2 B_2\left(\frac{m_0 }{L }\right) + \cdots\right] =
\alpha_0 Z_1\left(\alpha_0,\frac{m_0 }{L }\right).\label{68}
\end{gather}

Actually all coef\/f\/icients of these series tend to inf\/inity
in the limit   $L \rightarrow \infty $ because of the divergent
integrals. Therefore the concrete values    $\alpha_0$, $m_0 $ are
indef\/inite if usual renormalization scheme is used.

Analysis of   \cite{Brodsky} shows that in the result of the
renormalization the additional terms contained the parameter $\xi
$ in the expansion~(\ref{67}) are absent at least with an accuracy
$\sim 10^{-10}$ corresponding to  $L \lesssim 10^7$~GeV. Analogous
estimation is correct when calculating characteristics of other
electromagnetic processes in QED. It means that after passing to
the limit  $L \rightarrow \infty $  the formulas of type of
equation~(\ref{67}) give the practical algorithm for calculation
of any observed values in QED with very high accuracy. For that
purpose the internal inconsistency of the theory that is
conditioned by indetermination of the series~(\ref{68}) becomes
unessential.

We believe that the variational approach presented in the paper
allows one to preserve the computing possibilities of QED but to
avoid its logical inconsistency. As it was shown above, in this
approach the ``physical'' electron characteristics are def\/ined
by the quasi-particle collective excitation of the ``primary''
electron-positron f\/ield itself and are not conditioned by
additional parameters or bond with any hypothetical particle.
Therefore the perturbation  series that def\/ine the corrections
to the form factor conditioned by the interaction with the
transverse electromagnetic f\/ield are dif\/fered from
equation~(\ref{66}) qualitatively. According to
equation~(\ref{Eqn63}) for the vertex function in this case the
expansion parameter  is the value  $\alpha_0^{-1} \sim \alpha$ and
the form factor depends on the invariant parameters as follows:
\begin{gather*}
F (k) \equiv F \left(\alpha_0; \frac{k^2}{m^2}; \frac{m }{m_0
}\right), \qquad \alpha_0 \sim \alpha^{-1}, \qquad m_0 \sim m
/\alpha,
\end{gather*}
and the ef\/fective cut-of\/f momentum for the divergent integrals
is the value   $L_0 \sim m_0 \sim m/\alpha$.

Then the   following series arises instead of the expansion
(\ref{66}):
\begin{gather}
 F (k) \approx 1 +  \alpha^{-1}_0 \left[ F^{(0)}_{1} \left(0; \frac{k^2}{m^2};0\right) + \frac{m  }{m_0 } F^{(1)}_{ 1}
\left(0; \frac{k^2}{m^2};0\right)  + \cdots\right]  \nonumber\\
\phantom{F (k) \approx}{} + \alpha^{-2}_0 \left[ F^{(0)}_{ 2}
\left(0; \frac{k^2}{m^2};0\right) + \frac{m }{m_0 } F^{(1)}_{ 2}
\left(0; \frac{k^2}{m ^2};0\right) + \cdots\right] +
\cdots.\label{70}
\end{gather}

In order to calculate the observed value for the form factor
equation~(\ref{70}) should be completed by the series that express
the ``primary'' parameter   $\alpha_0$, $m_0 $ over the observed
values $\alpha$, $m$. These series have the following general
form:
\begin{gather}
 \alpha \approx \alpha^{-1}_0 [ \lambda_0 + \alpha^{-1}_0 \lambda_1 +  \alpha^{-2}_0 \lambda_2  +
\cdots ]= \alpha_0
\tilde{Z}(\alpha_0 ), \nonumber\\
m \approx m_0 \alpha^{-1}_0[\mu_0 + \alpha^{-1}_0 \mu_1 +
\alpha^{-2}_0 \mu_2  + \cdots] = m_0 \tilde{Z_1}(\alpha_0
).\label{71}
\end{gather}
According to equation~(\ref{Eqn64}) all coef\/f\/icients in these
series are f\/inite and and do not depend on additional
parameters.

Substitution of equation (\ref{71}) in equation (\ref{70}) shows
that the electron form factor calculated by means of the
considered approach contains only the f\/inite values and does not
depend on the additional parameters as it should be in accordance
with the analysis in~\cite{Brodsky}:
\begin{gather*}
\tilde{F}_{\rm th}(k) \approx 1 +  \alpha \tilde{F}^{(0)}_{  1}
\left(  \frac{k^2}{m^2} \right) + \alpha^2 \tilde{
F}^{(0)}_{  2} \left(  \frac{k^2}{m ^2} \right)+ \cdots , \nonumber\\
\tilde{F}^{(0)}_{  1} \left(  \frac{k^2}{m^2} \right) = \frac{1}{\lambda_0} F^{(0)}_{1} \left(0; \frac{k^2}{m^2};0\right),\nonumber\\
\tilde{F}^{(0)}_{  2} \left(  \frac{k^2}{m^2} \right) =
\frac{1}{\lambda_0^2}\left[\frac{\lambda_1}{\lambda_0
}F^{(0)}_{1} \left(0; \frac{k^2}{m^2};0\right) + \frac{1}{ \mu_0 }
F^{(1)}_{ 1} \left(0; \frac{k^2}{m^2};0\right) + F^{(0)}_{ 2}
\left(0;
\frac{k^2}{m^2};0\right)\right].
\end{gather*}

The considered qualitative analysis of the ``physical'' electron
form factor can be applied not only for the interaction with the
vacuum transversal electromagnetic f\/ield but also for any kind
of the external f\/ields. It means, for example, that after
renormalization with the self-localized electron states the QED
corrections for the ``physical'' electron bound in the Coulomb
f\/ield of the nucleus in the hydrogen-like atom should be
described by the same formulas that can be found with usual
renormalization scheme~\cite{Lamb}.

Actually it means that the considered form of QED  that uses the
basis of the self-localized states and the perturbation theory on
the parameter $\alpha^{ -1 }_0$ can lead to the same results for
the observed values as the renormalized series in the physical
f\/ine structure constant $\alpha$. They are dif\/fered only when
calculating the connections between the primary and observed
values of $\alpha_0$ and  $\alpha$ as well as $m_0$ and $m$.
However, the strict proof of the quantitative coincidence of our
results with known ones are connected with rather cumbersome
calculations of high-order corrections in a power of the parameter
$\alpha^{ -1 }_0$ on the basis of the vertex function
(\ref{Eqn63}) and the propagation function for the ``physical''
electron and should be considered separately.

In any case it is evident that for the practical purposes the
previously existing calculation scheme in QED is more
ef\/f\/icient than one  considered in the present paper. Therefore
the proof of the equivalence of the both forms of QED is important
mostly from the point of the internal self-consistency of QED.
Nevertheless, the suggested approach may have the applied interest
as well because it allows one to consider other leptons  as
excitations of the ``primary'' electron-positron f\/ield (see
Section~\ref{sec3} for $\mu$-meson). It is also may be essential
for the constructive use of the renormalization group in QED and
for the nonperturbative analysis of other quantum f\/ield models
including non-renormalized ones.

\section{Conclusions}\label{sec6}

Thus, in the present paper the self-localized quasi-particle
excitation of the electron-positron f\/ield   is found for the
f\/irst time in the framework of a standard form of the quantum
electrodynamics. Existence of a new state in the spectrum of  the
well investigated Hamiltonian is of great theoretical interest by
itself.

Besides, the physical interpretation of the obtained solution as
the ``physical'' electron has been considered. It allowed one to
f\/ind in the f\/inite form the relation between the charge $e_0$
and the mass $m_0$ of the ``bare'' electron considered as the
``primary'' parameters of the theory and observed values of the
charge $e_0$ and the mass $m_0$ of the ``physical'' electron. The
obtained relations between these values ($e_0  \sim e^{-1}$, $m_0
\sim m e^{-2}$)  show  that they cannot be calculated by means of
the standard perturbation theory.

The important consequence of the considered interpretation of the
self-localized state is the possibility to represent the Dirac
idea and to consider $\mu$-meson as an additional quasi-particle
excitation of the electron-positron f\/ield. It is shown that the
calculated mass of $\mu$-meson is very closed to its experimental
value.

It is also shown that the self-localized quasi-particle state does
not contradict to the Lorentz-invariance of the system and its
binding energy can be considered as the observed mass of this
excitation. It permitted us to describe the interaction of the
``physical'' electron with the transverse electromagnetic f\/ield
and to show that it can considered on the basis of the
perturbation theory as the series in a power of the observed
charge of electron  $e \sim e_0^{-1} \ll 1$. It proved that the
coef\/f\/icients of these series are free from the ultraviolet
divergence.

The internal structure of the ``physical'' electron considered as
the collective excitation does not depend on any additional
parameters with the exception of $\alpha$ and $m$. It is
described by the f\/inite functions without introduction of the
cut-of\/f momentum  $L$. It is also discussed quantitatively that
an existence of such structure does not contradict to the
experimental data about the electron form factor.

\subsection*{Acknowledgements}

Authors are grateful to ICTP Of\/f\/ice of External Activities for
the f\/inancial support of the participation in the International
Conference ``Symmetry in Nonlinear Mathematical Physics'' (June
24--30, 2007, Kyiv, Ukraine).

\pdfbookmark[1]{References}{ref}
\LastPageEnding

\end{document}